\pdfoutput=1
\documentclass[twocolumn,trackchanges]{aastex631}

\usepackage{float}

\usepackage{amsmath}
\usepackage{multirow}
\usepackage{makecell}
\usepackage{graphicx}

\begin{document}

\title{The Stellar Disk Structure Rrevealed by the Mono-age Populations of the LAMOST Red Clump Sample}

\correspondingauthor{Bingqiu Chen, Xiaowei Liu}
\email{bchen@ynu.edu.cn, x.liu@ynu.edu.cn}

\author{Zheng Yu}
\affiliation{South-Western Institute for Astronomy Research, Yunnan University, Kunming, Yunnan 650091, China}

\author[0000-0003-2472-4903]{Bingqiu Chen}
\affiliation{South-Western Institute for Astronomy Research, Yunnan University, Kunming, Yunnan 650091, China}

\author[0000-0001-5258-1466]{Jianhui Lian}
\affiliation{South-Western Institute for Astronomy Research, Yunnan University, Kunming, Yunnan 650091, China}

\author[0000-0002-8713-2334]{Chun Wang}
\affiliation{Tianjin Astrophysics Center, Tianjing Normal University, Tianjin 300387, China}

\author{Xiaowei Liu}
\affiliation{South-Western Institute for Astronomy Research, Yunnan University, Kunming, Yunnan 650091, China}

\begin{abstract}
Understanding the structure of the Galactic disk is crucial for understanding the formation and evolutionary history of the Milky Way. This study examines the structure of the Galactic disk by analyzing a sample of 138,667 primary red clump (RC) stars from the LAMOST and Gaia datasets. We have categorized these RC stars into mono-age populations and investigated their spatial distributions within the $R-Z$ plane, estimating scale heights and lengths through the fitting of their vertical and radial density profiles. Our analysis indicates that the vertical profiles of these mono-age populations fit a dual-component disk model, where both components exhibit significant flaring, particularly in the outer disk regions. Within a constant Galactocentric radius $R$, the scale heights of the first component, representing the morphologically thin disk, rise with age. In contrast, the scale heights of the second component, corresponding to the morphologically thick disk, remain comparatively stable across different age groups. Additionally, the radial density profiles of both disk components predominantly peak within a radial range of 7.5$-$8.5\,kpc. These findings underscore the importance of age as a crucial factor in shaping the spatial distribution and structural evolution of the Galactic disk, offering valuable insights into its complex dynamics and history. 
\end{abstract}


\section{Introduction} \label{sec:intro}

Our position within the Milky Way offers a unique internal viewpoint, allowing for detailed observations that provide stringent constraints on models of disk galaxy formation. The stellar disk, comprising approximately three-quarters of the Galactic stars \citep{rix2013milky}, is a crucial component of the Milky Way. It serves as a foundation for stellar archaeology, given that the spatial and age distributions of its stars offer valuable insights into the Galactic history. Mapping the stellar number density across the Milky Way disk and understanding its variation among stellar populations with different ages or chemical compositions is central to the field of Galactic astronomy. This research encompasses the characterization of the Galactic disk structure, its star formation, assembly, and perturbation history \citep{bovy2016stellar, ted2017age, xiang2018stellar, yu2021mapping}. Extensive data collection facilitated by large spectroscopic and photometric surveys such as the Apache Point Observatory Galactic Evolution Experiment (APOGEE; \citealt{majewski2017apache}, Gaia \citep{vallenari2023gaia}, the Large Sky Area Multi-Object Fiber Spectroscopic Telescope (LAMOST; \citealt{cui2012large}), the GALactic Archaeology with HERMES (GALAH; \citealt{martell2016galah}), and the Dark Energy Spectroscopic Instrument (DESI; \citealt{aghamousa2016desi}) has significantly contributed to our understanding of the positions, chemical abundances, ages, and kinematics of a vast array of disk stars.

In-depth analysis of the Galactic disk has confirmed that the Galaxy disk generally follows exponential laws in stellar density distribution. Previous works \citep{yoshii1982density, gilmore1983new, robin2003synthetic, juric2008milky, chen2017constraining} have demonstrated a geometric dichotomy between the Galactic thin and thick disks, discernible through star counts in the vertical direction. Variations in disk scale heights ($h_Z$)  and lengths ($h_R$) reported in the literature can often be attributed to differences in age-related selection effects across studies \citep{chang2011information, amores2017evolution}. Moreover, two distinct stellar populations have been identified in chemical abundance space, separated into high-[$\alpha$/Fe] and low-[$\alpha$/Fe] sequences in the [$\alpha$/Fe]–[Fe/H] plane \citep{fuhrmann1998nearby, fuhrmann2008nearby, prochaska2000galactic, bensby2003elemental, bensby2004possible, bensby2014exploring, reddy2006elemental, adibekyan2011new, adibekyan2012chemical, adibekyan2013kinematics, haywood2013age, bovy2016stellar, yu2021mapping}. Typically, the thick-disk populations are characterized by being more metal-poor and $\alpha$-enhanced, with older ages and hotter kinematics compared to the thin-disk populations \citep{adibekyan2013kinematics, bensby2014exploring}. A study by \citet{bovy2016stellar} using 14,699 red clump stars (RCs) from the APOGEE-RC catalog confirmed a structural dichotomy in both chemical and spatial distributions, with high-[$\alpha$/Fe] abundance patterns (mono-abundance populations) exhibiting a larger scale height and shorter scale length than their low-[$\alpha$/Fe] counterparts, alongside evidence of flaring in the latter. Similarly, \citet{yu2021mapping}, analyzing 96,201 primary red clump stars from LAMOST, found that both low- and high-[$\alpha$/Fe] mono-abundance populations show distinct spatial and kinematical properties, with notable exponential flaring in the outer disk regions ($R > 8$\,kpc).

Galactic disks are not only distinguished by their two-component structure but also exhibit intricate substructures such as warps and flares. These features have been extensively studied using various tracers, including H~{\sc i} gas \citep{levine2006vertical, koo2017tracing}, infrared emission from dust and cold giants \citep{freudenreich1994dirbe}, Cepheids \citep{chen2019intuitive, skowron2019mapping}, and stellar populations such as young OB stars and old Red Giant Branch (RGB) stars \citep{romero2019gaia}.

The eighth data release (DR8) of the low-resolution spectroscopic survey by LAMOST in March 2021 includes 11,214,076 optical spectra (spanning wavelengths from 3700 to 9000\,\AA) with a resolution of R $\sim$ 1800. More than 90\% of these spectra are of stellar objects. Utilizing methodologies similar to those of \citet{ting2018large}, \citet{wang2023precise} accurately distinguished between RGB and red clump (RC) stars in the LAMOST DR8 dataset \citep{luo2015first}. They employed a neural network approach with LAMOST–Kepler common stars as a training set, enabling precise calculations of age and mass for these stellar populations. The RGB and RC samples demonstrated remarkable purity and completeness, exceeding 95\% and 90\% respectively, with typical uncertainties in stellar mass and age of 9\% and 24\%, respectively.

In this study, we utilize the RC sample provided by \citet{wang2023precise} to explore the spatial structure of the Galactic stellar disk using a mono-age approach. The layout of this paper is as follows: In Section \ref{sec:Data}, we detail the RC sample data. Section \ref{sec:method} outlines our methodology for constructing number density maps for each mono-age population. In Section \ref{sec:results}, we present our results, including the fitting of stellar number density distributions and calculations of scale heights and scale lengths for each population. Section \ref{sec:discuss} is dedicated to discussions of these findings, and Section \ref{sec:conclusion} provides a summary of our study.

\section{Data} \label{sec:Data}

RCs are invaluable as standard candles for probing the three-dimensional structure of the entire Galactic disk due to their relatively consistent luminosity. Our study utilizes the RC sample from LAMOST as detailed by \citet{wang2023precise}. This sample effectively distinguishes between RGB and RC stars, the latter of which are further divided into primary and secondary RC stars based on mass distinctions \citep{bovy2014apogee, huang2020mapping}. This classification allows for precise age and mass estimations using a neural network approach, with the LAMOST–Kepler dataset serving as the training set \citep{wang2023precise}.

The stellar parameters for our analysis are sourced from the value-added catalog of LAMOST DR8 \citep{wang2022value}, which includes atmospheric properties such as effective temperature ($T_{\text{eff}}$), surface gravity (log\,$g$), and metallicity measurements. The catalog provides both [Fe/H], which represents iron abundance derived from iron lines, and [M/H], which is a measure of overall metallicity. Additionally, the catalog provides ratios of chemical abundances such as alpha-elements to metals ([$\alpha$/M]), carbon to iron ([C/Fe]), and nitrogen to iron ([N/Fe]), along with absolute magnitudes. It is noteworthy that for stars with spectral signal-to-noise ratios (SNR) above 50, the measurement precisions are as follows: $T_{\text{eff}}$ to within 85\,K, log\,$g$ to within 0.10\,dex, [Fe/H] and [M/H] to within 0.05\,dex, [C/Fe] and [N/Fe] to within 0.05\,dex and 0.08\,dex respectively, and [$\alpha$/M] to within 0.03\,dex. While both [Fe/H] and [M/H] are available, we primarily use [Fe/H] in our subsequent analysis due to its direct measurement from iron lines and its widespread use in stellar population studies. The specific application of [Fe/H] in our fitting procedures will be detailed in Section \ref{sec:RC distance}.

\section{Methodology} \label{sec:method}

\subsection{The Distance of RCs} \label{sec:RC distance}

Accurate distance estimation of RCs is crucial for assessing the structure of the Galactic disk. We derived distances for our RC sample by calibrating their absolute magnitudes in the $K_{\rm S}$ band, which is known to be a reliable distance indicator \citep{salaris2002population, laney2012new, girardi2016multiwavelength}. Despite the inherent accuracy, a systematic uncertainty of 3$-$5\% persists, primarily due to metallicity variations among these stars \citep{huang2020mapping}. In the current work, we have recalibrated the $K_S$ absolute magnitude for RCs considering metallicity effects, following the methodology of \citet{huang2020mapping}. To ensure both precision and a robust sample size, we apply selection criteria similar to those used in \citet{huang2020mapping}:
\begin{itemize}
    \item Galactic latitude $|b| \geq 30^\circ$,
    \item Reddening values $E(B-V) < 0.075$\,mag from \citet{schlegel1998maps},
    \item LAMOST spectra with a signal-to-noise ratio (SNR) greater than 30,
    \item Gaia Parallax relative uncertainties less than 15\%,
    \item 2MASS $K_S$ band photometric errors below 0.03\,mag.
\end{itemize}
Distances are calculated using the Gaia DR3 parallaxes \citep{lindegren2018gaia} and the intrinsic $K_S$ magnitudes are sourced from 2MASS, corrected for reddening using $E(B-V)$ values provided by \citet{schlegel1998maps}. The SNR threshold of 30 was chosen to optimize the balance between sample size and data quality---our analysis shows that at SNR$>$30, we achieve median uncertainties of 30\,K for $T_{\text{eff}}$, 0.042\,dex for log\,$g$, and 0.024\,dex for [Fe/H]. These uncertainties remain essentially unchanged when increasing to SNR$>$50 (30\,K, 0.041\,dex, and 0.024\,dex respectively), indicating that SNR$>$30 provides sufficient precision while maximizing the sample size. The resultant dataset comprised 9176 RCs. To accommodate metallicity effects, we fitted the $K_S$ band absolute magnitude $M_{K_S}$ as a function of [Fe/H], using a third-order polynomial:
\begin{equation}
    M_{K_S} = -1.59 - 0.097\,{\rm [Fe/H]} + 0.257\,{\rm [Fe/H]}^2 + 0.106\,{\rm [Fe/H]}^3.
\end{equation}

\begin{figure}[htbp]
	\centering
	\includegraphics[width=1.1\linewidth]{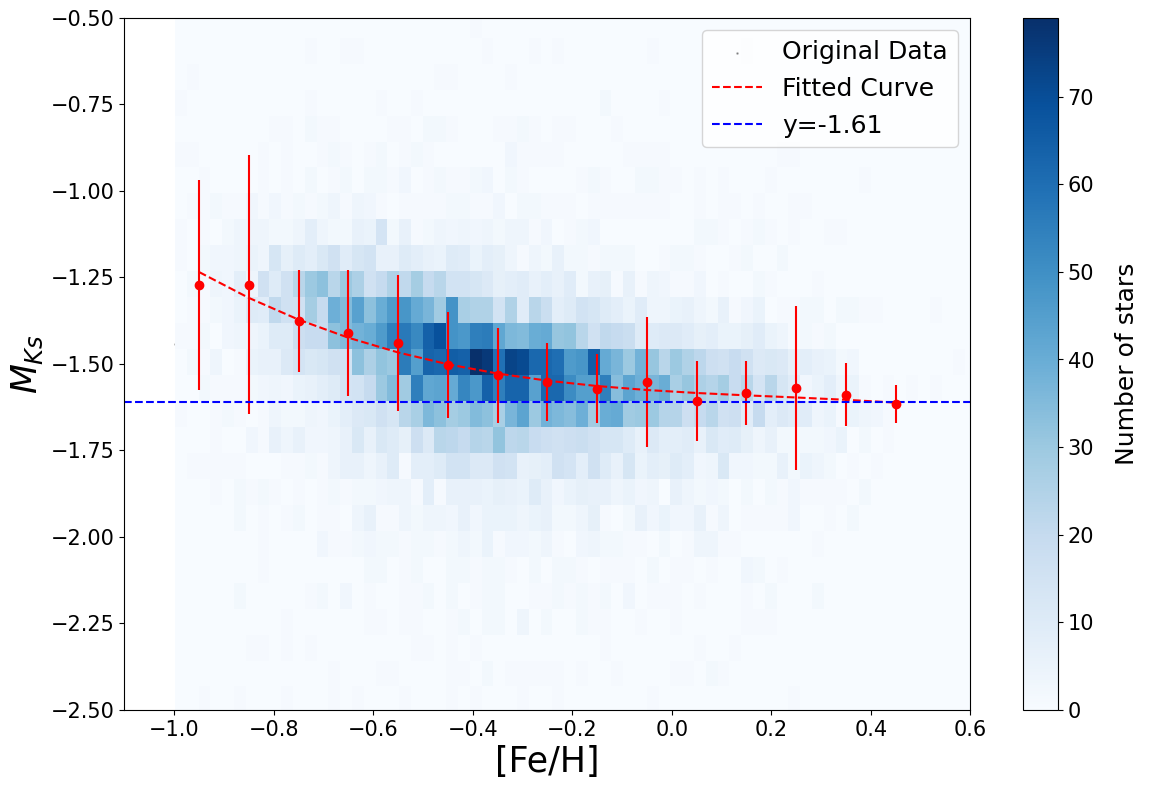}
	\caption{$K_S$ band absolute magnitudes of primary RCs as a function of metallicity [Fe/H]. Red points denote median values obtained by binning the data points into 15 bins with a bin size of 0.1\,dex in [Fe/H]. The error bars indicate the standard deviations of the individual bins. The red dashed line is a third-order polynomial fit to the red points.}
	\label{Fe_Mk}
\end{figure}

The results of the fitting, as shown in Fig.~\ref{Fe_Mk}, reveal that metal-poor RCs ([Fe/H] $< -$0.2\,dex) are significantly fainter by 0.1 to 0.4\,mag compared to their metal-rich counterparts ([Fe/H] $\geq-$0.2\,dex), which display an almost constant $M_{K_S}$ value around $-$1.61 mag, aligning with the findings of \citet{laney2012new}.

Furthermore, accurate distance calculation requires precise reddening correction. Following the approach of \citet{wang2019optical}, we fit the intrinsic color index $(J-K_S)_{0}$ as a function of stellar parameters ($T_{\text{eff}}$, [M/H], and log $g$):
\begin{align}
    (J - K_S)_0 = & -4.791 \log(T_{\text{eff}})^2 + 32.094 \log(T_{\text{eff}}) \nonumber \\
                & - 0.053 {\rm [Fe/H]}^2 - 0.033 {\rm [Fe/H]} \nonumber \\
                & + 0.057 \log\,g - 52.745.
\end{align}
The color excess $E(J-K_S) = (J-K_S)_{\text{obs}} - (J-K_S)_{\text{0}}$ is determined. Extinction value in the $K_S$ band is calculated using
$ A_{K_S} = \frac{R_{K_S}}{R_J - R_{K_S}} \cdot E(J-K_S)$,
where $R_{K_S}$ and $R_J$ are the extinction ratios for the $K_S$ and $J$ bands, respectively, adopted from \citet{yuan2013empirical}.
Finally, distances to RCs are computed using the standard relation $
    d = 10^{0.2(m_{K_S} - M_{K_S} - A_{K_S} + 5)}$.

\begin{figure}
    \centering
    \includegraphics[width=0.48\textwidth]{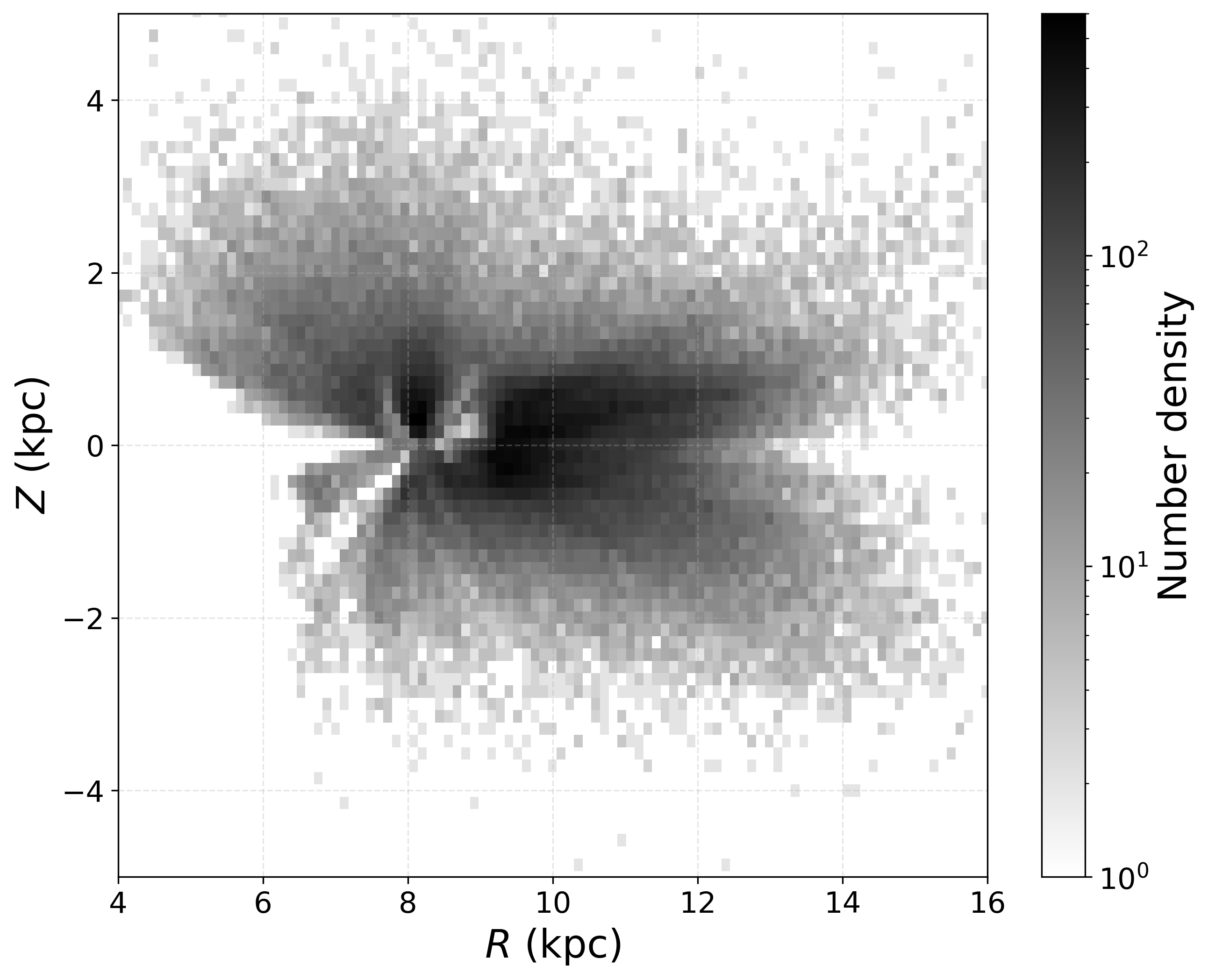}%
    \hfill%
    \includegraphics[width=0.48\textwidth]{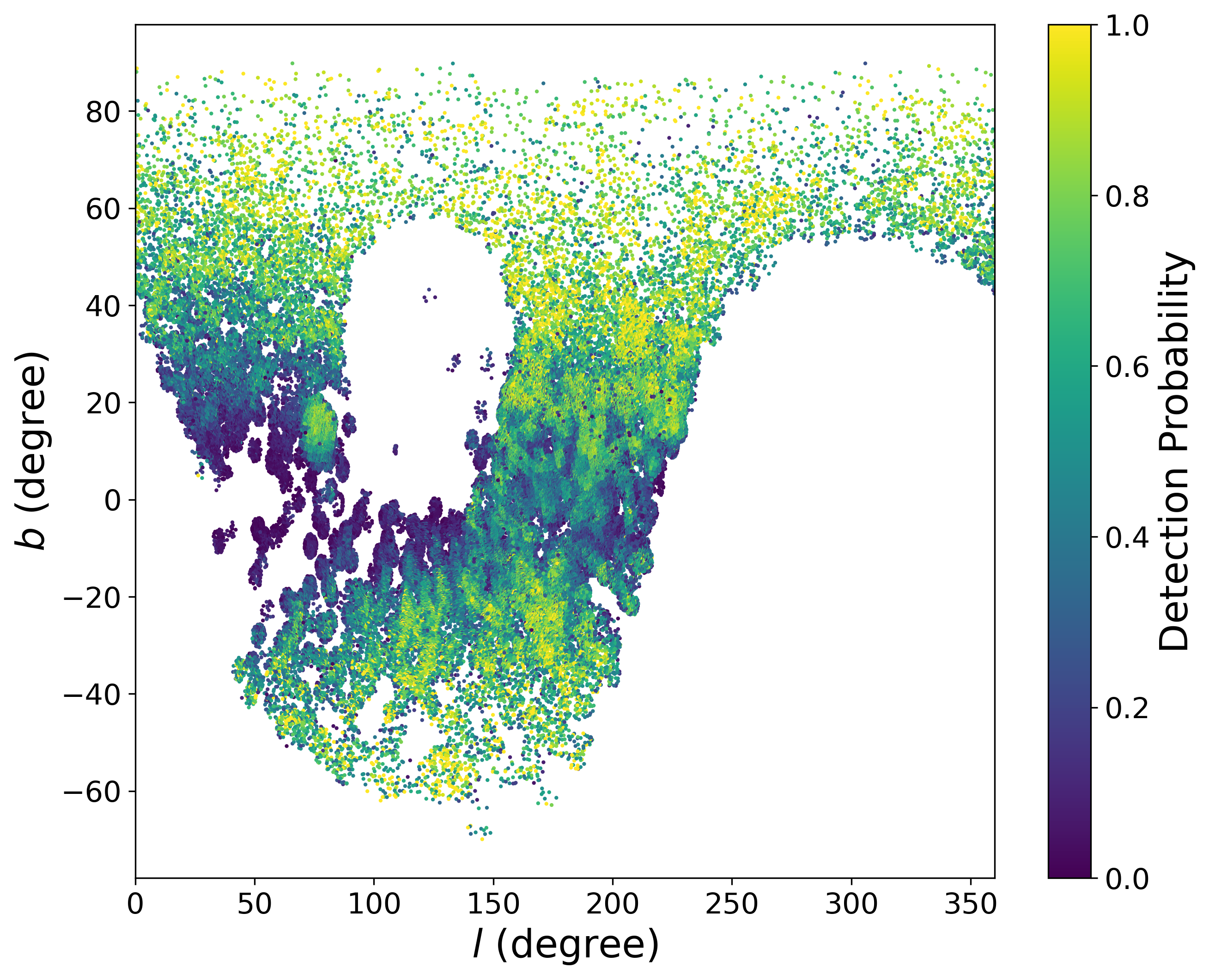}
    \caption{Top panel: The number density distribution of the $RC$ sample in $R-Z$ plane. The density map uses a bin size of 0.03 $\times$ 0.025\,kpc. The Sun is located at ($R$, $Z$) = (8, 0.025)\,kpc. Bottom panel: Sky maps of the selection function for our $RC$ sample, illustrating variations in detection probability across different Galactic coordinates. The scatter plot displays data points spanning Galactic latitudes from $-$60 to 80\,deg. The accompanying color bar indicates the detection probability, scaled from 0.0 (dark purple) to 1.0 (bright green).}
    \label{fig1}
\end{figure}

\subsection{Correction for Sample Selection Effects}\label{sec:Selection Func}

To ensure the integrity and validity of our findings on the structure of the Galactic disk as inferred from the spatial distributions of our RC sample, it is imperative to address the selection effects inherent in our dataset. Selection effects describe the probability of including a star from the parent catalog into our subsample, influenced by the completeness of the parent catalog and the specific filtering criteria applied \citep{chen2018selection}. Our RC sample originates from the LAMOST DR8 dataset. In the current work, we adopt the complete photometric data from Gaia DR3 to correct for the selection biases identified in the LAMOST DR8. 

We utilize the methodology proposed by \citet{castro2023estimating} to estimate the selection function applicable to our sample. This method employs Bayesian statistics and models the selection process as a binomial distribution with a uniform prior. We categorize both the parent Gaia DR3 catalog and the selected RC subsample according to sky position (HEALPix), $G$-band magnitude, and the color index ($G - G_{\rm RP}$). Within each category, we document the number of stars in the parent catalog ($n$) and in the subsample ($k$), then calculate the subsample membership probability ($p$) for each category. This probability is then combined with the parent catalog's selection function to derive the comprehensive subsample selection function. For these computations, we utilize the {\sc GaiaUnlimited} Python package\footnote{\url{https://github.com/gaia-unlimited/gaiaunlimited}}, which facilitates the generation of detailed sky maps and numerical tables representing the subsample selection function. Fig.~\ref{fig1} illustrates two key aspects of our RC sample. The top panel shows the observed number density distribution of RC stars in the $R-Z$ plane before selection function correction, with stars primarily distributed within 4-16\,kpc in Galactic radius and $\pm$4 kpc in height above and below the Galactic plane. The bottom panel showcases the sky maps of the derived selection function for our RC sample. This visual representation provides a detailed overview of the spatial distribution of our RC sample, with the color gradient from dark purple to bright green visually encoding the detection probability.

\begin{table*}
\centering
\caption{Range of parameters used for the MCMC fitting.}
\label{tab:mcmc}
\begin{tabular}{cccc}
\hline
\hline
Parameter & Range & Description \\
\hline
$h_{Z1}$ (kpc) & [0, 1.5] & Scale height of Component 1 \\
$h_{Z2}$ (kpc) & [0, 4] & Scale height of Component 2 \\
$\rho_1$ & [0, 3e5] & Midplane number density at $R_i$ \\
$f$ & [0, 0.5] & Fractional contribution of the second component \\
\hline
\end{tabular}
\end{table*}

\begin{table*}
\centering
\caption{Fitting results of the vertical density distribution with a double exponential model.}
\label{tab:all hz}
\begin{tabular}{cccccccr}
\hline
\hline
Age (Gyr) & $h_{Z1}$ (kpc) & $h_{Z2}$ (kpc) & $\ln(\rho_1)$ & $f$ & BIC$_\text{double}$ & BIC$_\text{single}$ \\
\hline

& & & $6<R<7$\,kpc & & & \\
\hline

0$-$3  &  0.198$\pm$0.056  &  1.115$\pm$0.364  &   $11.17 \pm 0.57$  &   0.047$\pm$0.031  & $ -9.380 $ & $ -0.247 $ \\
3$-$5  &  0.198$\pm$0.04  &  1.249$\pm$0.36  &    $11.51 \pm 0.48$  &   0.031$\pm$0.016  & $ -32.050 $ & $ 11.258 $ \\
5$-$7  &  0.274$\pm$0.054  &  1.415$\pm$0.311  &   $10.99 \pm 0.39$  &   0.042$\pm$0.018  & $ -50.125 $ & $ -2.003 $ \\
7$-$9  &  0.317$\pm$0.04  &  1.187$\pm$0.175   &   $10.63 \pm 0.10$  &   0.093$\pm$0.035  & $ -65.559 $ & $ 33.332 $ \\
$>$ 9  &  0.423$\pm$0.087  &  1.467$\pm$0.67  &   $10.26 \pm 0.28$  &   0.078$\pm$0.063  & $ -64.977 $ & $ -49.044 $ \\

\hline

& & & $7<R<8$\,kpc & & & \\

\hline
0$-$3  &  0.153$\pm$0.032  &  0.827$\pm$0.068  &   $11.99 \pm 0.34$  &   0.066$\pm$0.024  & $ -35.936 $ & $ 15.705$ \\
3$-$5  &  0.199$\pm$0.042  &  0.848$\pm$0.109  &   $11.76 \pm 0.34$  &   0.071$\pm$0.028  & $ -46.070 $ & $ 2.102$ \\
5$-$7  &  0.201$\pm$0.032  &  0.997$\pm$0.108  &   $11.89 \pm 0.30$  &   0.049$\pm$0.015  & $ -57.090 $ & $ 31.849$ \\
7$-$9  &  0.273$\pm$0.051  &  1.063$\pm$0.171   &  $11.38 \pm 0.31$  &   0.067$\pm$0.026  & $ -66.723 $ & $ -7.995$ \\
$>$ 9  &  0.284$\pm$0.058  &  1.01$\pm$0.128  &   $11.18 \pm 0.31$  &   0.096$\pm$0.032  & $ -71.905 $ & $ -26.399$ \\
\hline

& & &  $8<R<9$\,kpc & & &   \\

\hline

0$-$3  &  0.203$\pm$0.044  &  0.871$\pm$0.239  &   $11.58 \pm 0.26$  &   0.084$\pm$0.051  & $ -12.561$ & $ 13.142$ \\
3$-$5  &  0.207$\pm$0.032  &  0.854$\pm$0.116  &   $11.70 \pm 0.23$  &   0.08$\pm$0.032  & $ -37.977$ & $ 7.977$ \\
5$-$7  &  0.256$\pm$0.047  &  1.118$\pm$0.139  &   $11.52 \pm 0.30$  &   0.057$\pm$0.019  & $ -60.630$ & $ 20.971$  \\
7$-$9  &  0.279$\pm$0.055  &  0.914$\pm$0.101  &   $11.16 \pm 0.27$  &   0.129$\pm$0.049  & $ -66.065$ & $ -27.569$ \\
$>$ 9  &  0.415$\pm$0.076  &  1.207$\pm$0.492  &   $10.66 \pm 0.20$  &   0.092$\pm$0.073  & $ -72.255$ & $ -49.844$ \\
\hline

& & &  $9<R<10$\,kpc & & &   \\

\hline

0$-$3  &  0.274$\pm$0.056  &  1.116$\pm$0.273  &   $10.35 \pm 0.16$  &   0.146$\pm$0.084  & $ -30.544$ & $ 1.985$ \\
3$-$5  &  0.332$\pm$0.064  &  1.111$\pm$0.415  &   $10.30 \pm 0.15$  &   0.135$\pm$0.103  & $ -47.281$ & $ -9.350$ \\
5$-$7  &  0.433$\pm$0.064  &  1.556$\pm$0.713  &   $10.08 \pm 0.13$  &   0.078$\pm$0.069  & $ -64.888$ & $ -30.082$ \\
7$-$9  &  0.454$\pm$0.097  &  1.218$\pm$0.396  &   $9.86 \pm 0.15$   &   0.187$\pm$0.142  & $ -69.955$ & $ -54.081$ \\
$>$ 9  &  0.464$\pm$0.098  &  1.446$\pm$0.674  &   $9.80 \pm 0.17$   &   0.142$\pm$0.127  & $ -65.390$ & $ -53.541$ \\
\hline

& & &  $10<R<11$\,kpc & & &   \\
\hline
0$-$3  &  0.353$\pm$0.055  &  2.057$\pm$0.96  &   $9.13 \pm 0.16$  &   0.106$\pm$0.07  & $ -34.626$ & $ -19.665$ \\
3$-$5  &  0.462$\pm$0.062  &  2.109$\pm$0.99  &   $9.03 \pm 0.14$  &   0.09$\pm$0.08  & $ -50.380$ & $ -11.300$ \\
5$-$7  &  0.672$\pm$0.142  &  1.526$\pm$0.674 &   $8.46 \pm 0.16$  &   0.182$\pm$0.166  & $ -55.995$ & $ -50.961$ \\
7$-$9  &  0.67$\pm$0.131  &  2.365$\pm$0.787  &   $8.23 \pm 0.16$  &   0.189$\pm$0.125  & $ -64.642$ & $ -62.773$ \\
$>$ 9  &  0.648$\pm$0.136  &  2.384$\pm$0.738 &   $8.18 \pm 0.18$  &   0.212$\pm$0.125  & $ -74.339$ & $ -69.921$ \\
\hline

& & &  $11<R<12$\,kpc & & &   \\
\hline
0$-$3  &  0.386$\pm$0.069  &  2.04$\pm$0.998  &   $8.48 \pm 0.19$  &   0.102$\pm$0.079  & $ -37.348$ & $ -30.447$ \\
3$-$5  &  0.505$\pm$0.083  &   1.955$\pm$0.929  &    $8.31 \pm 0.17$  &   0.116$\pm$0.107  & $ -45.667$ & $ -21.192$ \\
5$-$7  &  0.619$\pm$0.101  &  2.753$\pm$0.807  &   $7.82 \pm 0.17$  &   0.158$\pm$0.086  & $ -48.786$ & $ -43.368$ \\
7$-$9  &  0.836$\pm$0.61  &  1.758$\pm$0.854   &   $7.39 \pm 0.38$  &   0.187$\pm$0.156  & $ 3.431$ & $ -51.112$ \\
$>$ 9  &  0.718$\pm$0.482  &  2.175$\pm$0.953  &   $7.38 \pm 0.39$  &   0.213$\pm$0.155  & $ -46.489$ & $ -61.297$ \\
\hline

& & &  $12<R<14$\,kpc & & &   \\
\hline
0$-$3  &  0.536$\pm$0.137  &  2.151$\pm$0.989  &   $7.63 \pm 0.26$  &   0.136$\pm$0.103  & $ -37.944$ & $ -38.867$ \\
3$-$5  &  0.755$\pm$0.161  &  1.733$\pm$0.923  &    $7.59 \pm 0.19$  &   0.146$\pm$0.146  & $ -39.963$ & $ -43.619$ \\
5$-$7  &  0.83$\pm$0.259  &  2.007$\pm$0.936  &   $7.29 \pm 0.23$  &   0.186$\pm$0.144  & $ -50.879$ & $ -58.558$ \\
7$-$9  &  0.833$\pm$0.563  &  2.686$\pm$0.969   &   $6.89 \pm 1.04$  &   0.239$\pm$0.136  & $ -54.011$ & $ -59.610$ \\
$>$ 9  &  0.704$\pm$0.192  &  2.934$\pm$0.679  &   $6.97 \pm 0.25$  &   0.273$\pm$0.109  & $ -53.171$ & $ -54.756$ \\
\hline
\end{tabular}
\end{table*}

\begin{figure*}
    \centering
    \includegraphics[width=\textwidth]{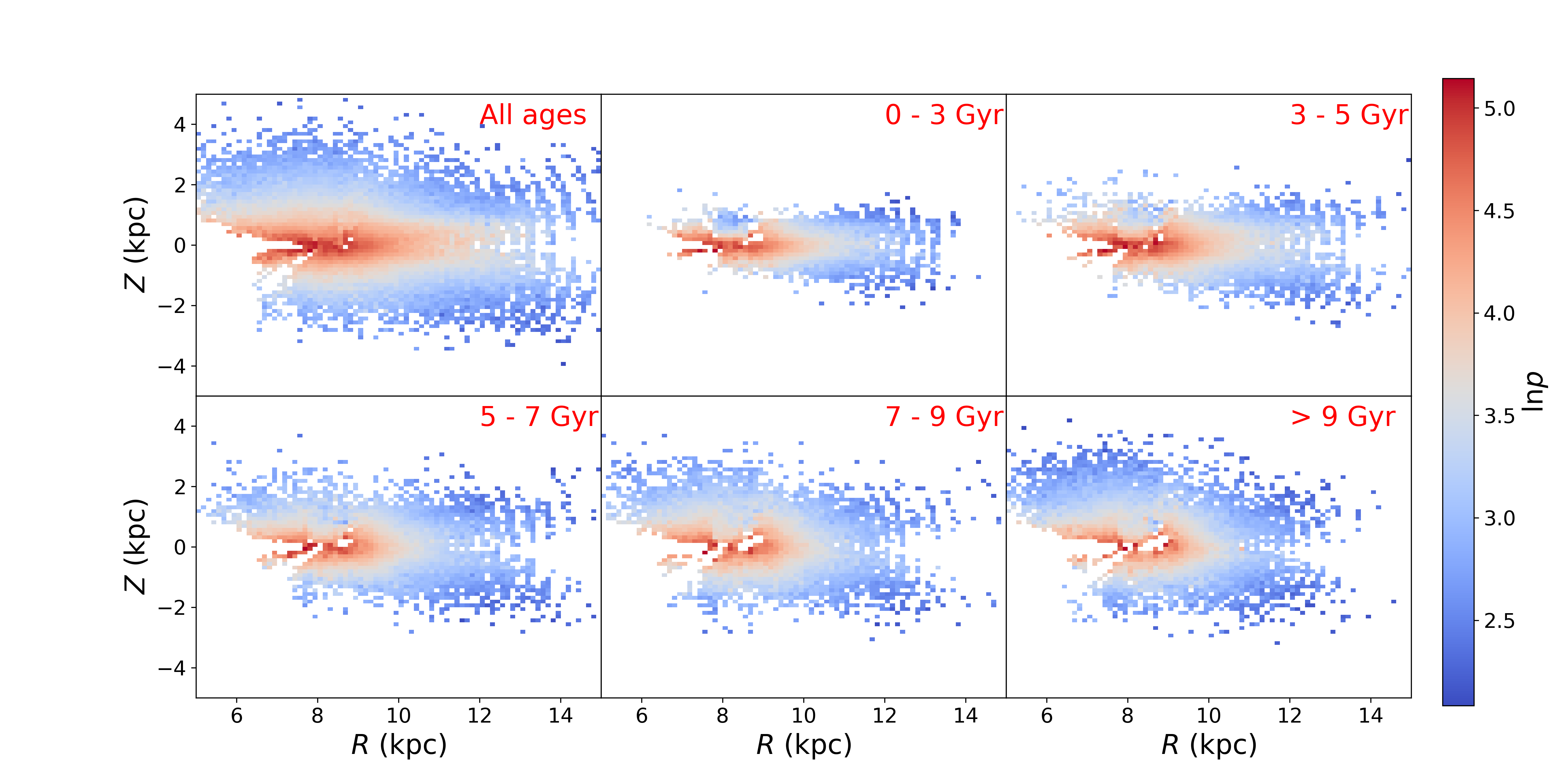}
    \caption{Distribution of stellar number density in the disk $R$–$Z$ plane for different mono-age populations, indicated in the upper right corner of each panel. The density map uses a bin size of 0.125 $\times$ 0.125\,kpc. Each panel represents a distinct age group, with density variations shown through a color gradient from blue (low density) to red (high density).}
    \label{fig2}
\end{figure*}

\begin{figure*}
    \centering
    \includegraphics[width=\textwidth]{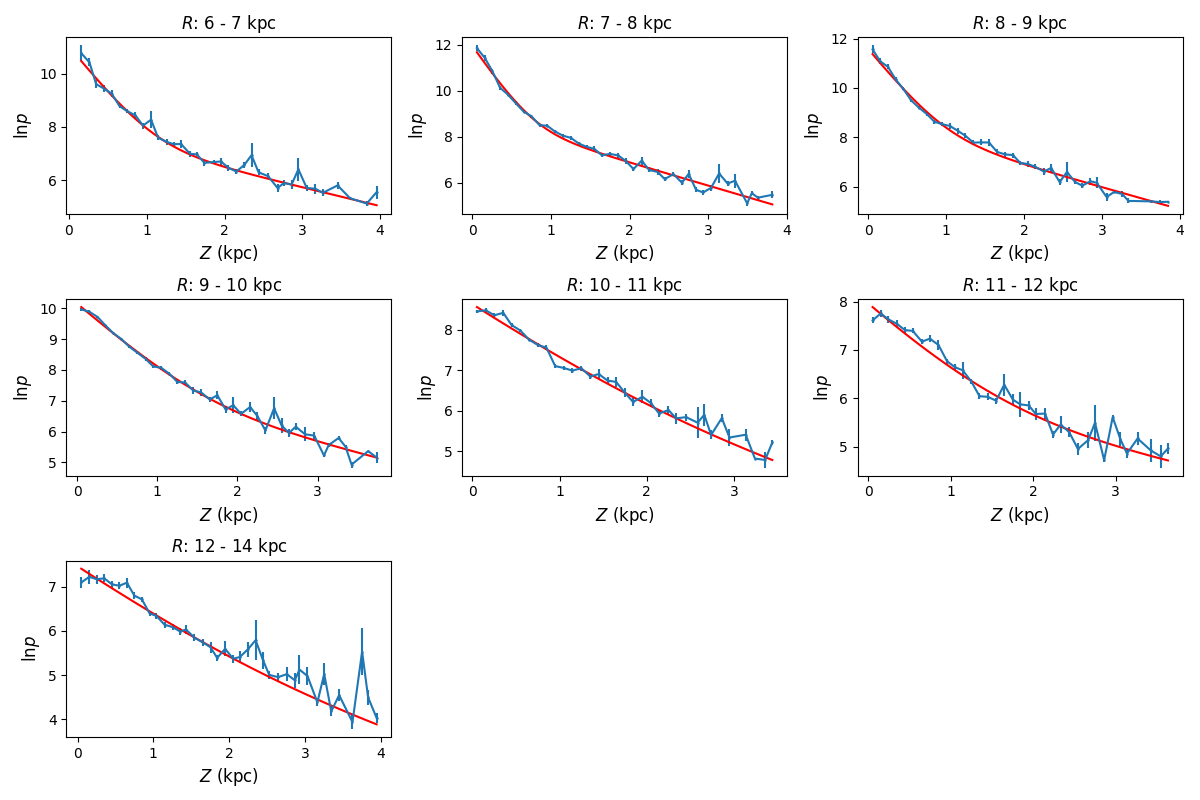}
    \caption{Fitting of vertical stellar number density for the mono-age population (5$-$7\,Gyr) using a double exponential function. Each panel shows results for a specific $R_i$ bin, marked at the top. Median values and standard errors in 100\,pc wide vertical bins ($Z_{\text{bins}}$) are depicted by blue lines and error bars, respectively. Red lines represent the best-fit curves.}
    \label{fig3}
\end{figure*}

\begin{figure*}
    \centering
    \includegraphics[width=0.6\linewidth]{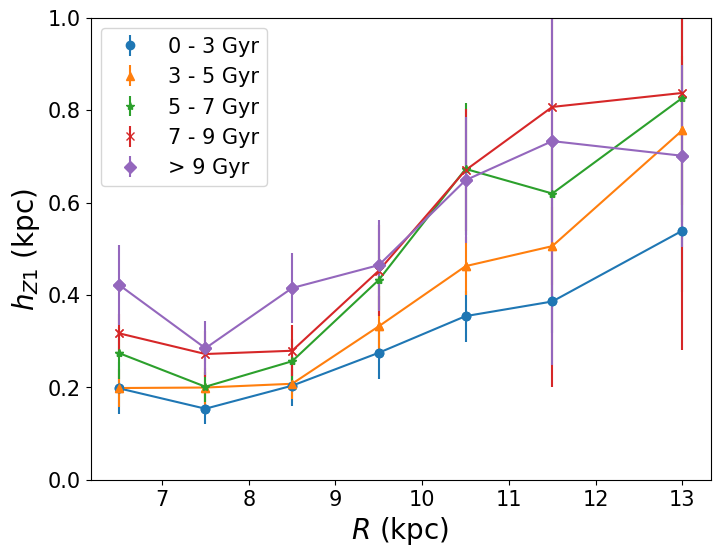}
    \includegraphics[width=0.6\linewidth]{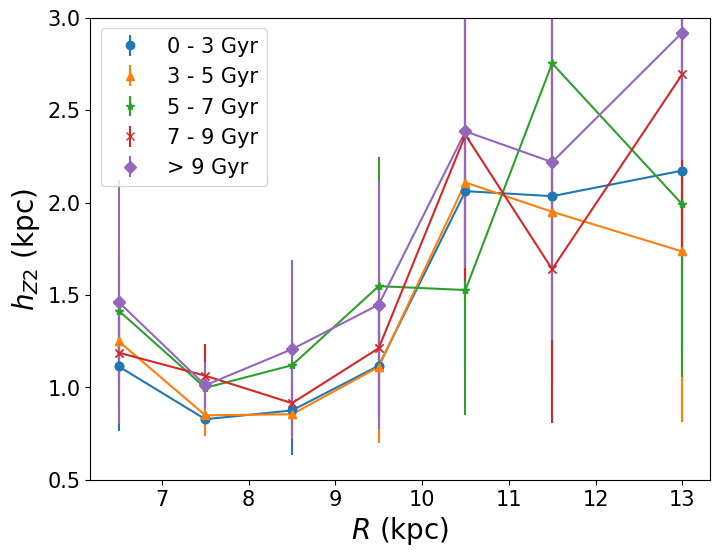}
    \caption{Vertical scale height profiles for various mono-age populations. Error bars indicate 1-sigma uncertainty. The top and bottom panels show the radial dependencies of $h_{Z1}$ and $h_{Z2}$, respectively, for different age groups, as indicated.}
    \label{fig4}
\end{figure*}

\begin{figure}
    \centering
    \includegraphics[width=0.5\textwidth]{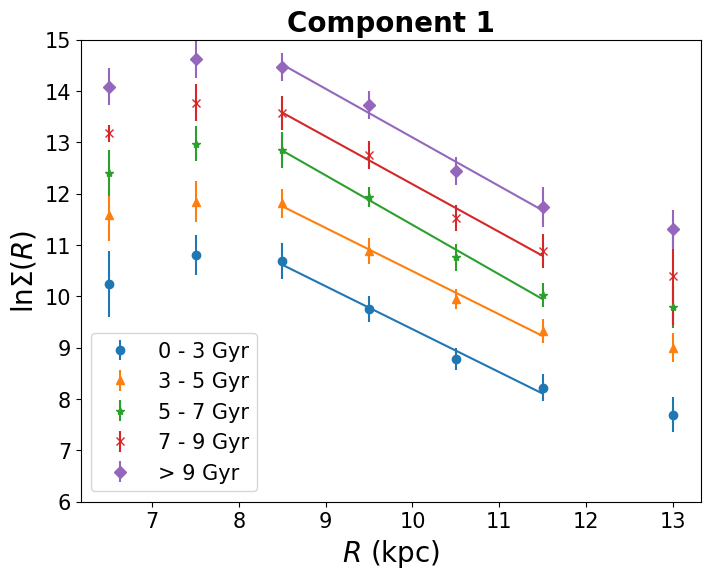}
    \includegraphics[width=0.5\textwidth]{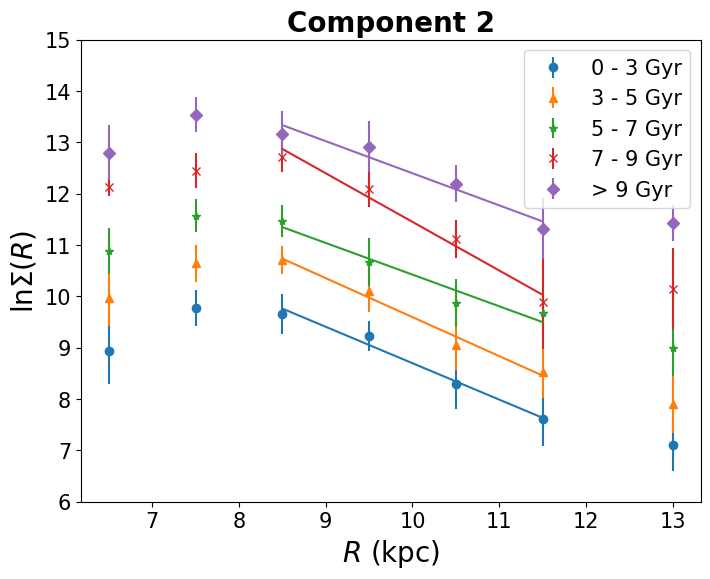}
    \caption{Radial surface density profiles $\Sigma(R)$ for the five mono-age populations. Different populations are offset vertically for clarity. The top panel highlights Component 1, while the bottom panel focuses on Component 2. Each point represents the natural logarithm of surface densities across different radial segments. The fitting results are shown as solid lines in the Figure.}
    \label{fig5}
\end{figure}

\subsection{Calculation of the Stellar Number Density}

To derive a three-dimensional distribution of stellar number density within the Galactic disk, we focus on mono-age populations of RC stars. Based on the ages derived from \citet{wang2023precise}, the RC sample is divided into five age-based subgroups: 0$-$3\,Gyr, 3$-$5\,Gyr, 5$-$7\,Gyr, 7$-$9\,Gyr, and $>$ 9\,Gyr. It is important to acknowledge that uncertainties in stellar age determinations can impact the classification of mono-age populations, potentially affecting subsequent model fitting. These uncertainties may lead to some degree of mixing between adjacent age bins, particularly for older stellar populations where relative age errors are typically larger. Consequently, when comparing our results to theoretical models, it is crucial to consider the effects of age uncertainties. We address this issue comprehensively in \hyperref[appendixA]{Appendix A}, where we demonstrate that our methodology remains robust in recovering age-dependent trends, despite the potential presence of the bin mixing. This analysis provides confidence in the reliability of our results while emphasizing the need for careful interpretation in the context of age uncertainties.

For each mono-age population, stars are grouped into bins based on the distance modulus, denoted by $\mu$, with a fixed bin size of 0.2\,mag. The stellar number density, $\rho(d_i)$, within a volume element $\Delta V(d_i)$ for a distance bin $d_i$, is obtained by:
\begin{equation}
\rho(d_i) = \frac{N(d_i)}{\Delta V(d_i)},
\label{eq:Hdi}
\end{equation}
where $N(d_i)$ represents the count of stars within the volume bin $\Delta V(d_i)$ at distance $d_i$, corrected for selection effects. The volume $\Delta V(d_i)$ is given by:
\begin{equation}
\Delta V(d_i) = \frac{\omega}{3} \left(\frac{\pi}{180}\right)^2 (d_{i2}^3 - d_{i1}^3),
\end{equation}
where $\omega$ is the area of the field in square degrees, and $d_{i1}$ and $d_{i2}$ are the lower and upper distance limits of each $d_i$ bin, respectively.

Fig.~\ref{fig2} illustrates the $R-Z$ distributions of stellar number densities for both the overall sample and each mono-age population. We have applied corrections for selection effects in calculating the number densities for individual $\mu$ bins, as detailed in Section \ref{sec:Selection Func}. These visualizations clearly depict how stellar number densities vary with age and spatial distribution within the Galactic disk. Notably, intermediate- and old-age populations ($>$ 5\,Gyr) are distributed over a broad range, from 6 to 14\,kpc in radius $R$ and from $-$4 to 4\,kpc in height $Z$. In contrast, younger populations (0$-$3\,Gyr) are predominantly found within more restricted regions, specifically from 6 to 14\,kpc in $R$ and $-$2 to 2\,kpc in $Z$. This pattern demonstrates significant variations in distribution correlated with the age of stellar populations.

\section{Results}\label{sec:results}

\subsection{Vertical Profile Fitting for Mono-age populations}\label{subsec:vertical_profile}

We analyze the vertical profiles of mono-age RC populations by categorizing them into different radial bins ($R$ bins). Previous studies have demonstrated that vertical density distributions are well-represented by exponential functions, using either single or double exponential models (e.g., \citealt{bovy2016stellar} for mono-abundance populations; \citealt{xiang2024formation} for mono-age populations). In this work, we find that the vertical number density distribution of each population is best modeled using a double exponential function:
\begin{equation}
\rho = \rho_{1} \left(\exp\left(-\frac{|Z - Z_0|}{h_{Z1}}\right) + f \exp\left(-\frac{|Z - Z_0|}{h_{Z2}}\right)\right),
\label{eq:hz12}
\end{equation}
where $\rho_1$ denotes the local stellar number density at the Galactic midplane, $Z_0$ the vertical position of the the Galactic midplane, set to 25\,pc \citep{juric2008milky}, $h_{Z1}$ and $h_{Z2}$ the scale heights for the two disk components, and $f$ the fractional contribution of the second component. 

The mono-age populations are divided into bins based on their projected Galactocentric distance $R_i$. For $R_i < 12$\,kpc, the bin size is 1\,kpc, and for $R_i > 12$\,kpc, it increases to 2\,kpc due to lower star counts. Each $R_i$ bin is assumed to have a constant scale height $h_Z(R_i)$. The double exponential model parameters for each $R_i$ bin across all mono-age populations are estimated using a Markov Chain Monte Carlo (MCMC) approach. This method optimizes the fit parameters within the ranges specified in Table~\ref{tab:mcmc}. The parameter ranges in Table~\ref{tab:mcmc} were empirically determined based on our experience with Galactic structure studies. These ranges are consistent with findings from recent literature (e.g., \citet{yu2021mapping} and \citet{xiang2018stellar}) while allowing for a broader exploration of the parameter space. The scale height ranges ($h_{Z1}$ and $h_{Z2}$) accommodate various stellar populations and potential variations across different Galactic regions. The $\rho_1$ range is set to cover expected stellar densities, and the $f$ limit assumes the first component is typically dominant. These empirically chosen ranges ensure a comprehensive parameter space exploration while maintaining physical plausibility. In our MCMC implementation, we used 500 walkers, each running for 40,000 steps, with 4 dimensions corresponding to the number of parameters being fit.

Fig.~\ref{fig3} illustrates the model fitting results across various radial bins $R_i$ for the intermediate age mono-age population (5$-$7\,Gyr). Within each $R_i$ bin, the stellar density data are segmented into vertical bins $Z_{\text{bins}}$ of 100\,pc each. We utilize the median values of these bins to fit the model, thereby minimizing the impact of outliers. Error bars in Fig.~\ref{fig3} denote the 1-sigma uncertainties, which increase in areas of higher $Z$ due to diminishing stellar counts.

The figure shows that the vertical number density in each radial bin is well described by a double exponential model. Detailed fitting results for additional mono-age populations are provided in \hyperref[appendixB]{Appendix B}. For all examined mono-age populations, the lower $Z$ regions predominantly exhibit the first component of the model, which aligns with the traditional characteristics of the Galactic thin disk, having smaller scale heights. Conversely, the second component, which emerges prominently at higher $Z$, is consistent with the properties of the traditional thick disk, as evidenced by its larger derived scale heights. These results are comprehensively tabulated in Table~\ref{tab:all hz}, detailing the vertical number density distributions.

To substantiate the validity of employing a double exponential model over a single exponential model for fitting, we introduced the Bayesian Information Criterion (BIC; \citealt{schwarz1978estimating}). The BIC is defined as:

\begin{equation}
\mathrm{BIC} \equiv -2 \ln \mathcal{L}_{\max} + k \ln N,
\label{eq:BIC}
\end{equation}

where $\mathcal{L}_{\max}$ represents the maximum likelihood value derived from the model, $k$ denotes the number of free parameters in the model, and $N$ is the number of data points fitted by the model. In Table~\ref{tab:all hz}, we present the BIC values for both models across each radial ($R$) interval for every mono-age population. As evident from Table~\ref{tab:all hz}, the double exponential model yields notably lower BIC values compared to the single exponential model in the majority of radial intervals. This indicates that the double exponential model provides a superior fit to the data points. However, it is noteworthy that in some instances, particularly for larger radial distances ($R >$ 12\,kpc) and older stellar populations (age $>$ 7\,Gyr and $R >$ 12\,kpc), the double exponential model produces slightly higher BIC values than the single exponential model. This discrepancy may be attributed to limitations in fitting arising from the paucity of distant and older stars in our sample.

Fig.~\ref{fig4} displays the variation of scale heights $h_{Z1}$ and $h_{Z2}$ as a function of Galactic radius $R$ across different mono-age populations. The upper panel of Fig.~\ref{fig4} depicts a trend in which $h_{Z1}$ typically reduces up to a radius of $R = 7.5$\,kpc, and subsequently increases beyond this radius for each age group. On the other hand, the lower panel indicates similar patterns for $h_{Z2}$. We note that $h_{Z1}$ displays a direct correlation with age, whereas $h_{Z2}$ does not follow a consistent age-related pattern. In general, for all age groups, the increase in the scale heights of Component 2 ($h_{Z2}$) in relation to $R$ is similar, displaying stronger trends when compared to Component 1. This observation aligns with the findings of \citet{lian2022milky}. Additionally, it is noteworthy that for Component 1, the rate of increase in scale heights in older populations is marginally faster than that in younger populations.

A particularly intriguing feature evident in Fig.~\ref{fig4} is the minimum in scale heights near the solar position ($R$ $\approx$ 7.5\,kpc) across all age groups. This characteristic aligns with findings from other studies, such as \citet[Fig.~6]{yu2021mapping}, suggesting it may be an intrinsic feature of the Galactic disk structure. Two key observations support this interpretation. Firstly, the consistency of this trend across all stellar age populations, including both young and old stars, indicates that it is unlikely to be a consequence of spiral arm dynamics, which primarily affect younger stellar populations. Secondly, the replication of this feature in studies employing different selection function corrections, such as \citet{yu2021mapping}, argues against it being an artifact of our specific bias-addressing methodology. These factors collectively suggest that the observed minimum in scale height near the solar radius is likely a genuine structural characteristic of the Milky Way's disk.

This feature becomes particularly interesting when considered alongside recent findings by \citet{lian2024broken}. In their Fig. 1, they report a broken shape in the luminosity surface density profiles near the solar neighborhood, with a transition point around 7.5\,kpc. Similarly, our Fig.~\ref{fig5} shows radial surface density profiles exhibiting a break near the solar position ($R$ $\approx$ 7.5\,kpc), where we observe both a minimum in scale heights and a maximum in scale lengths. While using different approaches - luminosity surface density in \citet{lian2024broken} and stellar surface density in our work - both studies independently reveal this structural transition near the solar radius. The consistency in finding these profile breaks through different methodological approaches suggests this is likely a genuine structural feature of the Galactic disk rather than an artifact of analysis methods. This apparent coincidence of structural transitions near the solar position warrants further investigation into its physical origin.

\subsection{Radial Profile Fitting for Mono-age Populations}\label{sec:4.2}

We then analyze of the radial profile of the mono-age RC populations. Building on the work by \citet{bovy2016stellar}, we understand that the radial surface density profile of low-[$\alpha$/Fe] mono-abundance populations deviates from a single exponential model, favoring a broken exponential profile. This broken profile peaks at a radius $R_{\text{peak}}$ and then declines, indicating a more complex distribution than previously understood due to the observational constraints that limited earlier studies.

To effectively model the radial distribution for each mono-age population, we adopt the exponential function as detailed by \citet{bovy2016stellar}, which is expressed as:
\begin{equation}
\ln \Sigma(R) \propto
    \left\{
        \begin{aligned}
            -h_{R,\text{in}}^{-1} (R-R_{0}) \;\;\;\; &\text{for } R \leq R_{\text{peak}}, \\
            -h_{R,\text{out}}^{-1} (R-R_{0}) \;\;\;\; &\text{for } R > R_{\text{peak}},
        \end{aligned}
    \right.
\label{eq:surface}
\end{equation}
where $R_0$ denotes the radial distance from the Sun to the Galactic center. Parameters $h_{R,\text{in}}$ and $h_{R,\text{out}}$ represent the scale lengths inside and outside of $R_{\text{peak}}$, respectively. The surface density $\Sigma(R)$ for each radial segment is calculated by:
\begin{equation}
\Sigma(R) = \int_{-\infty}^{+\infty} \rho(Z, R) \, dZ = 2\rho_1 h_Z(R),
\label{eq:hz1hz2}
\end{equation}
where $\rho_1$ is the local stellar number density at the Galactic midplane and $h_Z$ is the vertical scale height. In the current study, we focus our fitting efforts on the radial profile for $R > R_{\text{peak}}$ due to insufficient data points available for $R \leq R_{\text{peak}}$. Due to data limitations at extreme radii, the fitting are constrained to $R > R_{\text{peak}}$ and excludes outlier data points at the highest radii ($R = 13$\,kpc) to maintain model integrity. The results of this targeted analysis are illustrated in Fig.~\ref{fig5}. The $R_i$ bins are divided using the same methodology as applied in our analysis of vertical scale heights.

The top panel of Fig.~\ref{fig5} reveals a consistent pattern among the mono-age populations, with a gentle rise in density before $R_{\text{peak}}$ (set to 8\,kpc), followed by a steeper decline post-peak. This trend is consistent across both panels, which show the density profiles for all mono-age populations in Component 1 (top panel) and Component 2 (bottom panel).

\section{Discussion}\label{sec:discuss}

We have obtained the detailed structure of the Galactic disk as a function of stellar ages. In this section, we will first compare our results with previous works and then discuss our findings, emphasizing their implications on the structure and formation of the Galactic stellar disk.

\subsection{Comparison of Vertical Density Profiles with Previous Works}

Our study benefits from a more comprehensive spatial coverage of the LAMOST RC sample. In Section \ref{subsec:vertical_profile}, we have focused on deriving the vertical density profile of individual mono-age populations within the radial range from 6.5 to 13\,kpc. Here, we explore the vertical structures by comparing mono-age populations with mono-abundance populations. This comparison is based on the assumption that populations within narrow bins of [$\alpha$/Fe] and [Fe/H] represent stellar populations with varying age distributions.

\citet{bovy2016stellar} observed that their low-[$\alpha$/Fe] mono-abundance populations align well with an exponentially flaring scale height $h_Z$($R$), characterized by a gradual increase in scale height with Galactocentric radius ($R_{\text{flare}}^{-1}$ = $-$0.12 $\pm$ 0.01\,kpc$^{-1}$) in the APOGEE RC sample. \citet{yu2021mapping} confirmed this flaring trend in similar exponential profiles outward for low-[$\alpha$/Fe] mono-abundance populations. Conversely, \citet{bovy2016stellar} found that high-[$\alpha$/Fe] populations exhibit a consistent scale height ($R_{\text{flare}}^{-1}$ = 0.0 $\pm$ 0.02\,kpc$^{-1}$). However, \citet{yu2021mapping} identified an obvious flaring vertical profile for their high-[$\alpha$/Fe] mono-abundance populations, although less pronounced than for low-[$\alpha$/Fe] mono-abundance populations. \citet{ted2017age} also reported evidence of flaring in individual mono-abundance populations, with $R_{\text{flare}}^{-1}$ = $-$0.06 $\pm$ 0.02\,kpc$^{-1}$ for high-[$\alpha$/Fe] mono-abundance populations and $R_{\text{flare}}^{-1}$ = $-$0.12 $\pm$ 0.01\,kpc$^{-1}$ for low-[$\alpha$/Fe] mono-abundance populations.

Our findings are consistent with those of \citet{ted2017age} and \citet{yu2021mapping}. As illustrated in Fig.~\ref{fig4}, both young and old mono-age populations exhibit a clear flaring phenomenon, with an increase in disk thickness beyond 8\,kpc. \citet{minchev2017minchev} suggested that the superposition of mono-age populations within mono-abundance populations could lead to decreased flaring in high-[$\alpha$/Fe] populations, even though the individual mono-age populations themselves continue to flare. \citet{mateu2018galactic} proposed that flaring becomes more prominent beyond 11\,kpc, particularly in the outer regions of the disk. \citet{lian2022milky} further suggested that a piece-wise model could effectively capture the flaring behavior, especially in the outer disk beyond 10\,kpc.

Flaring in the Galactic disk can arise from various mechanisms, including external perturbations from satellite galaxies, stellar migration, and a tilt in the mass distribution of the Galactic halo \citep{garcia2021flaring, han2023tilted}. By estimating density profiles for mono-age populations, we can better constrain the formation mechanisms of flaring and its impact on the disk's structure and evolution.

The top panel (Component 1) of Fig.~\ref{fig4} shows that older populations ($>$ 5\,Gyr) exhibit larger scale heights and more pronounced flaring in the outer disk. This behavior is consistent with expectations if flaring is primarily driven by radial migration; as stars migrate outward, their vertical excursions increase \citep{minchev2012radial}. Interestingly, our findings diverge from those of \citet{ted2017age}, who observed an increasing trend of $R_{\text{flare}}^{-1}$ with age, suggesting that the youngest populations flare most. \citet{ted2017age} speculated this might be due to an old Milky Way population that has undergone mergers, thereby reducing flaring in the oldest populations. Structural discontinuities resulting from mergers \citep{martig2014dissecting} might also play a role, particularly if mergers reduce flaring caused by radial migration \citep{minchev2014chemodynamical}. Notably, \citet{ted2017age} acknowledged that age uncertainties, which can be substantial (up to 40\%), may artificially increase age bin sizes, leading to the superposition of populations with different scale heights and flaring trends. In contrast, our larger sample size (138,667 stars) and more precise age uncertainties (typically 24\%) mitigate this mixing of age bins. The bottom panel (Component 2) of Fig.~\ref{fig4} reveals flaring in the outer disk for all mono-age populations within the thick disk. However, these populations cannot be effectively separated by age. 

\begin{table}
    \centering
\caption{Scale lengths for each mono-age population in the outer disk.}
\label{tab:length}
\begin{tabular}{ccc}
\hline
\hline
Mono-age Populations & $h_{R1,\text{out}}$ (kpc) & $h_{R2,\text{out}}$ (kpc) \\
\hline
0$-$3\,Gyr & 1.19$\pm$0.10 & 1.40$\pm$0.13 \\
\hline
3$-$5\,Gyr & 1.18$\pm$0.07 & 1.31$\pm$0.12 \\
\hline
5$-$7\,Gyr & 1.03$\pm$0.06 & 1.62$\pm$0.27 \\
\hline
7$-$9\,Gyr & 1.08$\pm$0.09 & 1.08$\pm$0.11 \\
\hline
$>$ 9\,Gyr & 1.05$\pm$0.09 & 1.58$\pm$0.26 \\
\hline
\end{tabular}
\end{table}

\begin{figure*}
    \centering
    \begin{minipage}{0.45\linewidth}
        \centering
        \includegraphics[width=\linewidth]{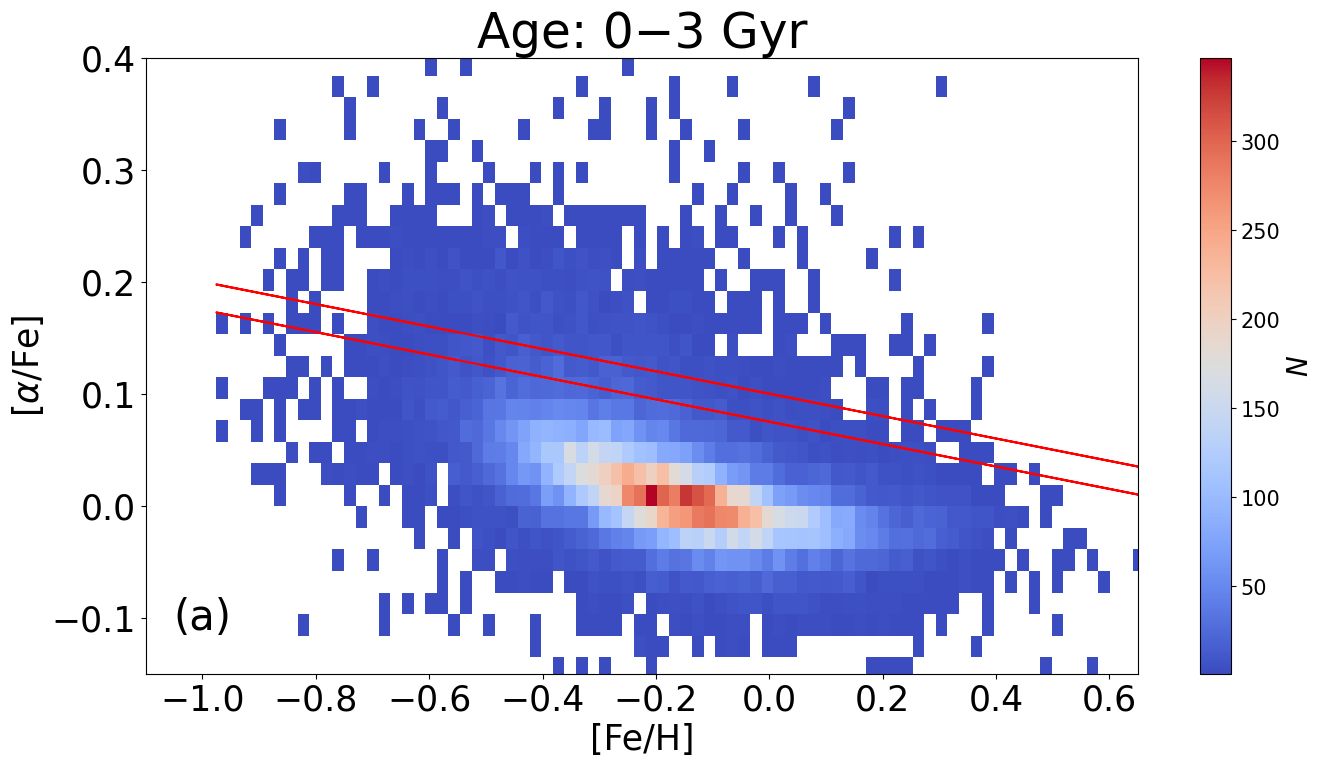}
    \end{minipage}
    \hfill
    \begin{minipage}{0.45\linewidth}
        \centering
        \includegraphics[width=\linewidth]{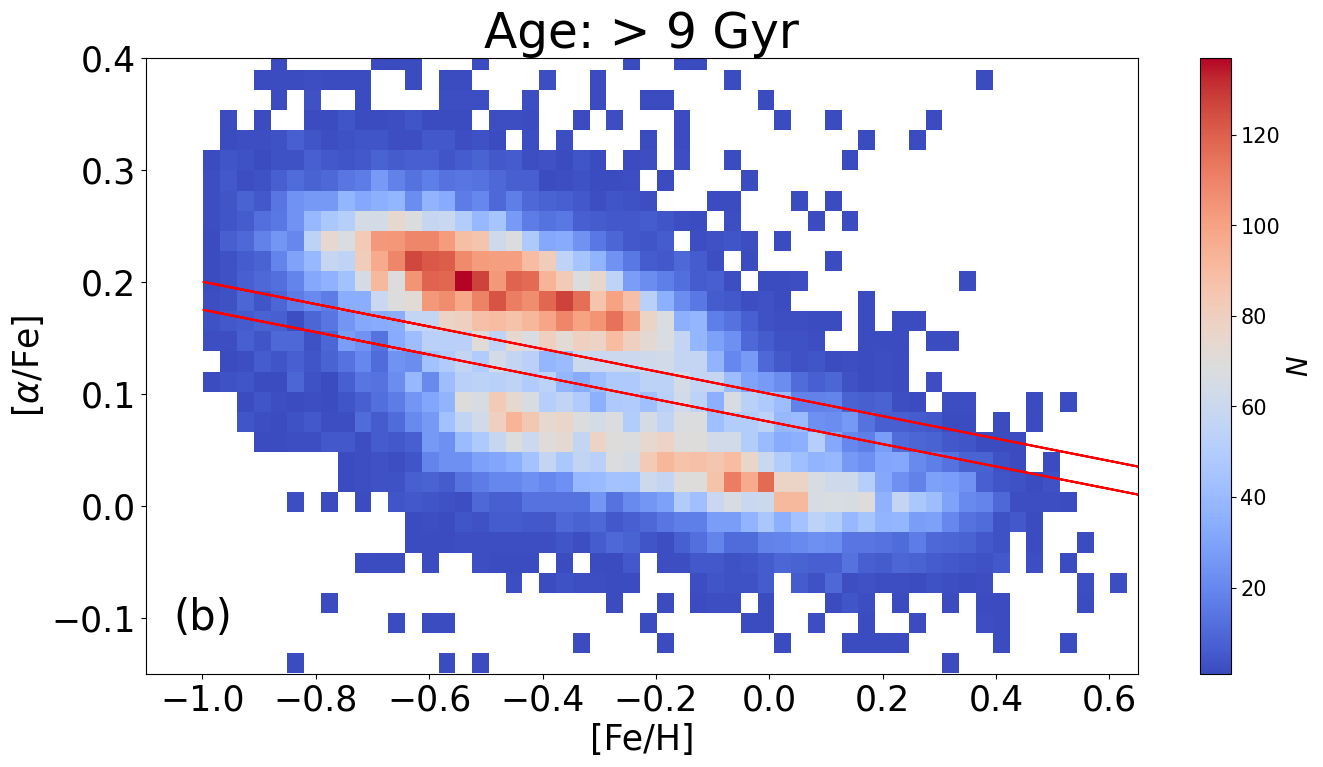}
    \end{minipage}
    \vspace{0.05cm} 
    \begin{minipage}{0.45\linewidth}
        \centering
        \includegraphics[width=\linewidth]{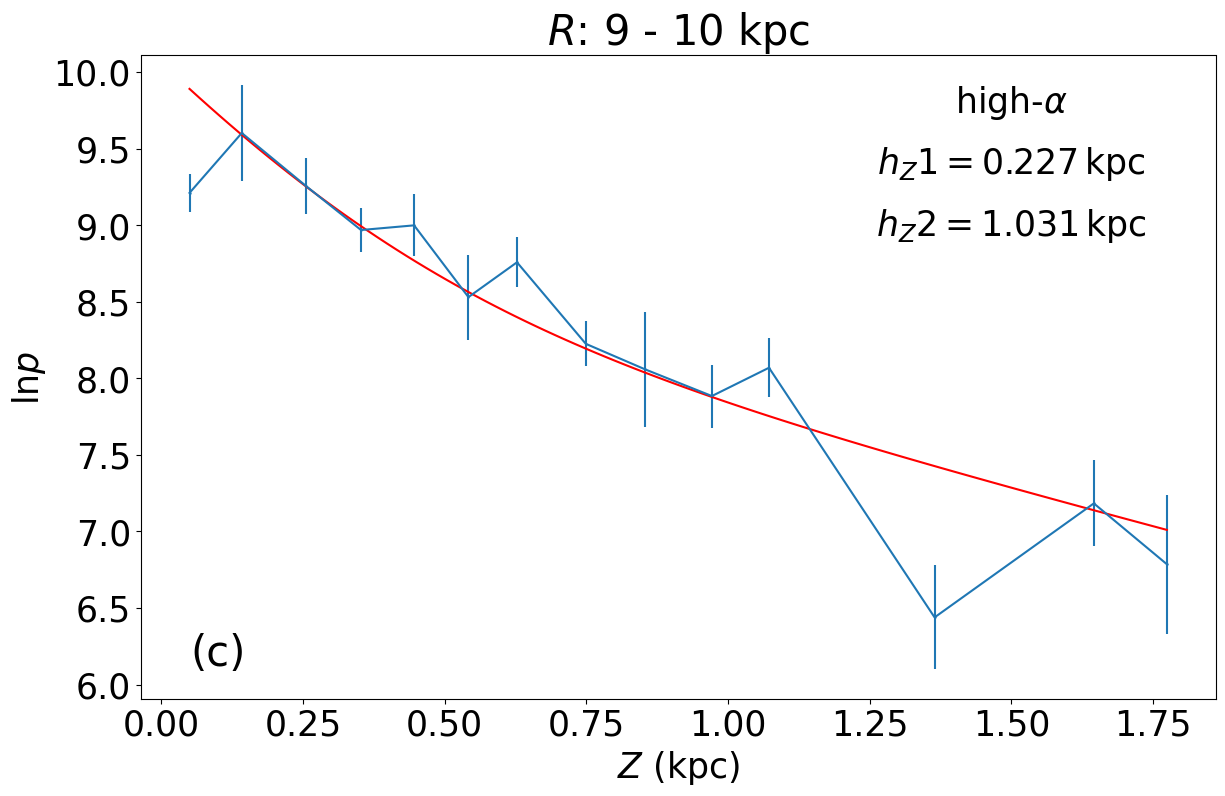}
    \end{minipage}
    \hfill
    \begin{minipage}{0.45\linewidth}
        \centering
        \includegraphics[width=\linewidth]{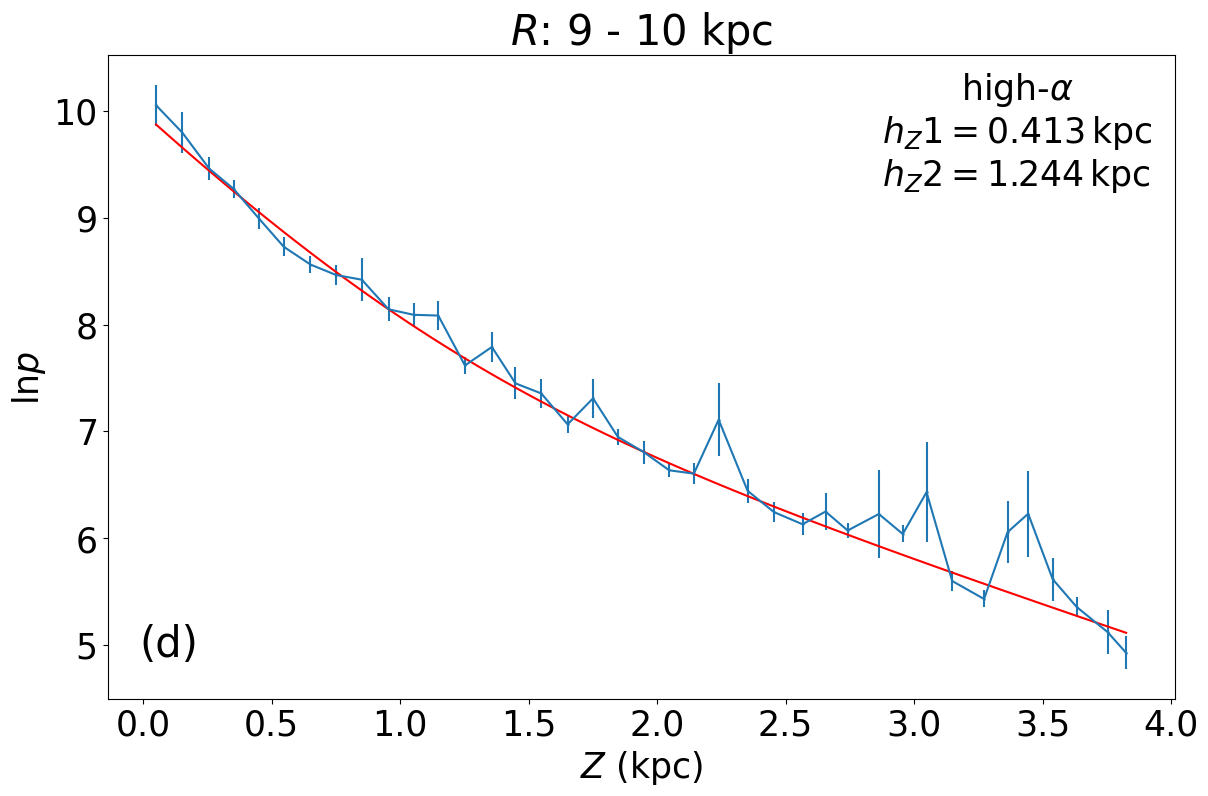}
    \end{minipage}
    \vspace{0.05cm} 
    \begin{minipage}{0.45\linewidth}
        \centering
        \includegraphics[width=\linewidth]{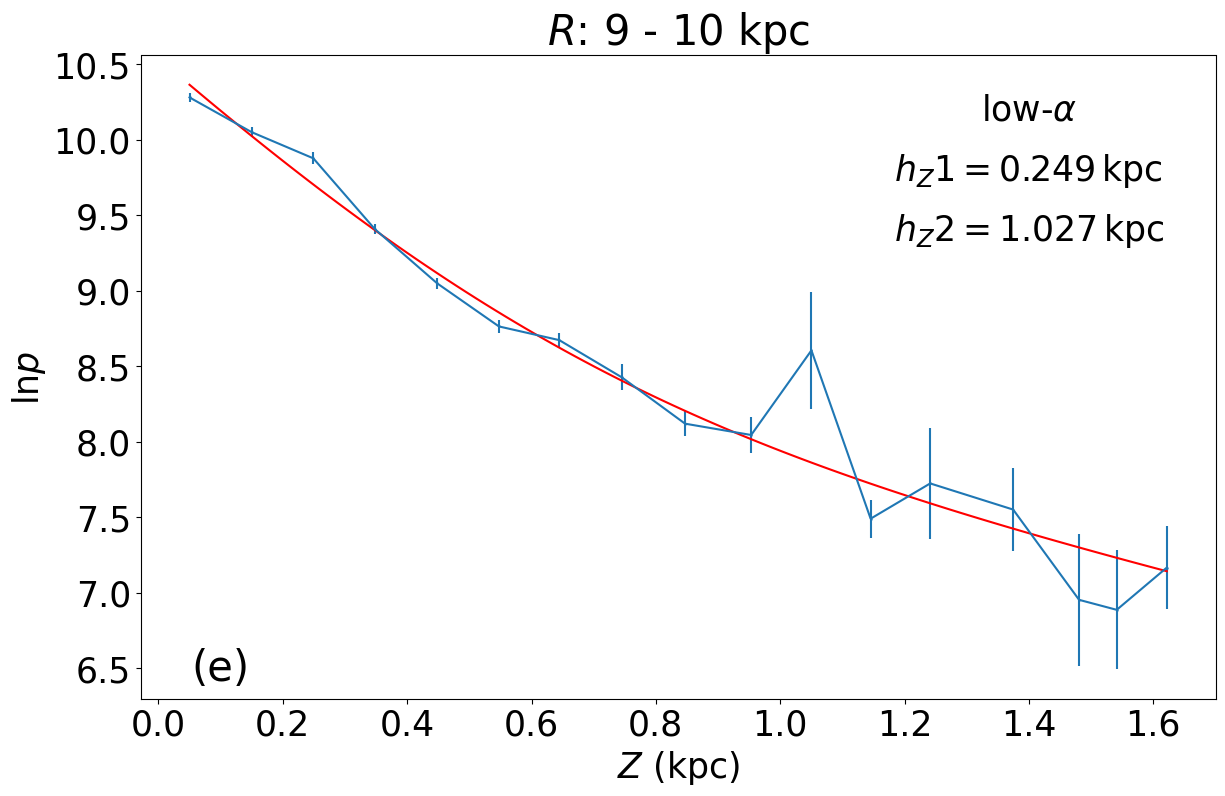}
    \end{minipage}
    \hfill
    \begin{minipage}{0.45\linewidth}
        \centering
        \includegraphics[width=\linewidth]{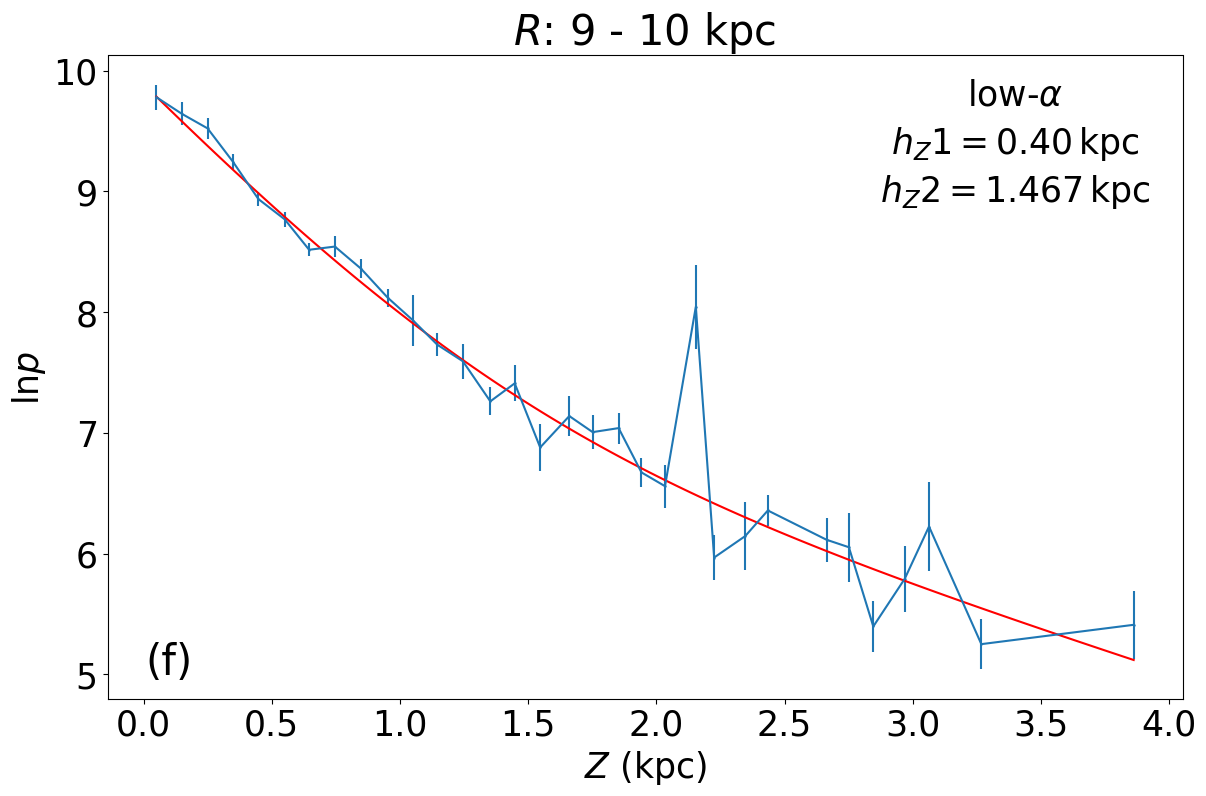}
    \end{minipage}
    \caption{[Fe/H]-[$\alpha$/Fe] distributions for our RC sample stars across two age populations: 0$-$3\,Gyr (left column) and $>$9\,Gyr (right column). The red demarcations in Panels (a) and (b) distinguish the chemical thin and thick disk stars. Panels (c) and (e) respectively depict double exponential fits for high- and low-$\alpha$ sequences of the 0$-$3\,Gyr population within the 9 $<$ $R$ $<$ 10\,kpc range. Similarly, Panels (d) and (f) respectively present double exponential fits for high- and low-$\alpha$ sequences of the $>$ 9\,Gyr population.}
    \label{fig6}
\end{figure*}

\subsection{Comparison of Radial Density Profiles with Previous Works}

As illustrated in Fig.~\ref{fig5}, radial surface density profiles are well represented by a broken exponential model, consistent with the findings of \citet{bovy2016stellar}. The break radii ($R_{\text{peak}}$) for our mono-age populations range from 7.5 to 8.5\,kpc. As for the scale length, Component 2 (morphologically thick disk) exhibits a broader profile compared to Component 1 (morphologically thin disk), as detailed in Table 3. These observations align with the study by \citet{ted2017age}, which reported a relatively constant surface-mass density weighted mean $R_{\text{peak}}$ for low-[$\alpha$/Fe] populations across different ages (we refer to the top panel of their Fig.~15). However, no similar trend was observed for their high-[$\alpha$/Fe] populations, suggesting distinct formation and evolutionary processes.
    
For the LAMOST RC sample, both components demonstrate a consistent $R_{\text{peak}}$ over time, as depicted in Fig.~\ref{fig5}. \citet{bovy2016stellar} observed variability in $R_{\text{peak}}$ across three low-[$\alpha$/Fe] mono-abundance populations, with larger values for lower [Fe/H]. This relationship was corroborated by \citet{ted2017age}, who showed a decreasing $R_{\text{peak}}$ as a function of [Fe/H]. However, our results do not reflect these observations. Our analysis reveals that the $R_{\text{peak}}$ for both components of the mono-age populations remain confined to a range of approximately 7.5 to 8.5\,kpc, suggesting that the relationship between metallicity and age is not a simple correlation. Furthermore, the dispersion in metallicity is more pronounced in older stellar populations\citep{xiang2022time, anders2023spectroscopic}.

Morphologically, we have determined differing scale lengths for Components 1 and 2 across various ages, as presented in Table~\ref{tab:length}. Contrary to previous studies which defined the traditional thin and thick disks based on chemical abundances ([$\alpha$/Fe] and [Fe/H]), we observed that Component 2 has a larger scale length than Component 1. This finding diverges from earlier research such as \citet{bensby2011first}, \citet{li2018formation}, and \citet{yu2021mapping}, which reported shorter scale lengths for the thick disk compared to the thin disk. These discrepancies may stem from the differences in the stellar populations sampled via chemical versus ages and morphological criteria.

\subsection{Chemical Properties within Mono-age Populations}

This section delves into the chemical characteristics of mono-age RC populations across the Galactic disc, focusing on the youngest (0$-$3\,Gyr) and oldest ($>$ 9\,Gyr) age groups. We categorize each mono-age population into high-$\alpha$ and low-$\alpha$ stellar groups based on an empirical division in the [Fe/H]-[$\alpha$/Fe] plane, as illustrated in Fig.~\ref{fig6}(a) and (b). While the bimodal distribution is more pronounced in the older population (Panel (b)), we apply a consistent classification approach to both age groups. The red demarcations in these panels represent our empirical division between chemical thin and thick disk stars. Following the methodology detailed in Section \ref{subsec:vertical_profile}, we have fitted their vertical density profiles using a double-exponential model, as depicted in Fig.~\ref{fig6}.

The analysis reveals a higher prevalence of stars with lower [$\alpha$/Fe] ratios in the younger RC populations, while older populations predominantly exhibit higher [$\alpha$/Fe] ratios. Notably, both the young and old RC populations show significant representation in both high- and low-[$\alpha$/Fe] sequences, indicating a bimodal distribution. Complementary to this, \citet{yu2021mapping} observed similar age distributions, with primary and secondary age peaks at 10 and 4.5\,Gyr, respectively, for high-[$\alpha$/Fe] mono-abundance populations. The low-[$\alpha$/Fe] populations typically ranged from 2$-$4\,Gyr, occasionally extending beyond 4\,Gyr.

The complexity of these number density distributions necessitated the adoption of a double exponential model for fitting, as single exponential models were inadequate. Each mono-age group displayed distinct scale heights for high-$\alpha$ and low-$\alpha$ populations ($h_{Z1} < 0.5$\,kpc; $h_{Z2} > 0.5$\,kpc), suggestive of dynamical perturbations influencing the chemical properties of the Galactic disk \citep{ted2017age}. Moreover, with a typical relative error of 24\% in age determinations, there is a substantial risk of misclassification of older stars into different mono-age categories. This issue highlights the critical need for more precise age measurements to enhance our understanding of the properties and dynamical history of the Galactic disk.

\section{Conclusion}\label{sec:conclusion}

This study investigates the structure of the Galactic disk by analyzing a dataset of 138,667 primary RC stars from the LAMOST and Gaia surveys. We have recalibrated the absolute magnitude in the $K_S$ band to improve distance accuracy and made adjustments for selection effects to accurately derive number density distributions of these RC stars. Our analysis characterizes the spatial distribution of mono-age populations across the $R-Z$ plane and quantifies the scale heights and lengths by fitting their vertical and radial density profiles. 

We find that the vertical profiles of mono-age populations are best described by a dual-component disk model, where both components exhibit pronounced flaring, especially in the outer regions of the disk. Notably, the flaring intensity of the first component, which corresponds to the morphologically thinner part of the disk, increases with age. This suggests that radial migration plays a significant role in shaping the disk structure. Moreover, within a constant Galactocentric radius $R$, the scale heights of the first component stars increase notably with age, while those of the second component, associated with the morphologically thicker disk, remain relatively stable across different age groups.Additionally, we observe that the radial density profiles for both two disk components predominantly peak within an $R$ range of 7.5$-$8.5\,kpc. The scale lengths of the two components vary with age, with the morphologically thicker disk having a larger scale length than the morphologically thinner disk.

Accurate age determinations, combined with detailed analyses of kinematics and chemical abundances, are vital for deepening our understanding of the Galactic formation and evolution. The observed complexities in the spatial and age distributions of these disk components underscore the need for further research. Future studies aimed at unraveling the underlying mechanisms responsible for these observed patterns will be crucial for enhancing our understanding of the structural dynamics of the Galactic disk.

\section*{Acknowledgments}

This work is partially supported by the National Key R\&D Program of China No. 2019YFA0405500, National Natural Science Foundation of China 12173034 and 12322304, the National Natural Science Foundation of Yunnan Province 202301AV070002 and the Xingdian talent support program of Yunnan Province. We acknowledge the science research grants from the China Manned Space Project with NO.\,CMS-CSST-2021-A09, CMS-CSST-2021-A08 and CMS-CSST-2021-B03. 

This work is based on data acquired through the Guoshoujing Telescope. The Guoshoujing Telescope (the Large Sky Area Multi-Object Fiber Spectroscopic Telescope, LAMOST) is a National Major Scientific Project built by the Chinese Academy of Sciences. Funding for the project has been provided by the National Development and Reform Commission. LAMOST is operated and managed by the National Astronomical Observatories, Chinese Academy of Sciences.

This publication makes use of data products from the European Space Agency (ESA) space mission Gaia. Gaia data are being processed by the Gaia Data Processing and Analysis Consortium (DPAC). Funding for the DPAC is provided by national institutions, in particular the institutions participating in the Gaia MultiLateral Agreement (MLA). The Gaia mission website is https://www.cosmos.esa.int/gaia. The Gaia archive website is https://archives.esac.esa.int/gaia.

\vspace{5mm}

\bibliography{mono-age}{}
\bibliographystyle{aasjournal}

\appendix

\section{Appendix: THE EFFECT OF UNCERTAINTIES ON TRENDS WITH AGE}
\label{appendixA}

To assess the impact of age uncertainties on our derived structural parameters, we conducted tests using mock data. We selected stars from our sample with Galactocentric radii between 9 and 10\,kpc, assuming their current age estimates as the "true" ages. We then created a mock dataset by introducing a 30\% random error to these ages, corresponding to the median age uncertainty in our sample. We modeled the vertical density distribution using a double exponential profile, with scale heights $h_{Z1}$ and $h_{Z2}$ for Component 1 and Component 2, respectively. The input model assumed both $h_{Z1}$ and $h_{Z2}$ increase monotonically with age across five bins (0$-$3, 3$-$5, 5$-$7, 7$-$9 and $>$9\,Gyr), with $h_{Z1}$ ranging from 0.2 to 0.6\,kpc and $h_{Z2}$ from 1.0 to 1.4\,kpc.

Crucially, after applying the 30\% random errors to the ages, we re-binned the mock sample into the same age bins as used in our original analysis. This re-binning step is essential, as it replicates the misclassification of stars into incorrect age bins due to age uncertainties, allowing us to directly assess the impact on the derived structural parameters for each mono-age population. We then applied our density fitting method to this re-binned mock sample, measuring the structural parameters for each apparent mono-age population. This approach enables us to quantify how age uncertainties affect not only individual stellar age estimates but also the composition and resulting structural parameters of our defined mono-age populations.

\begin{figure*}
    \centering
    \includegraphics[width=\textwidth]{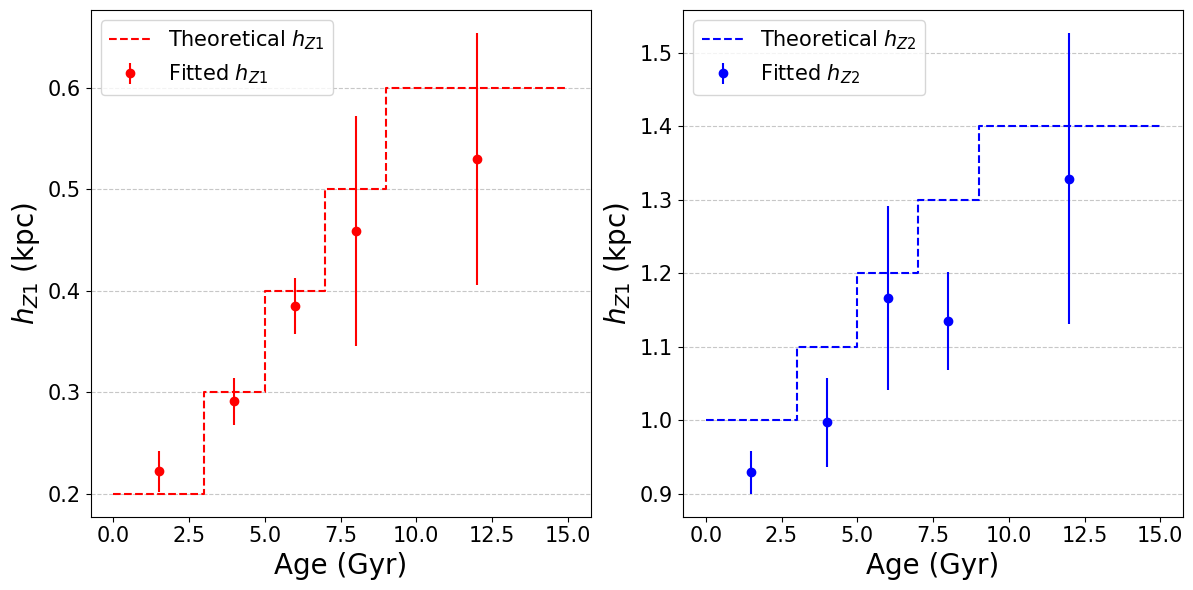}
    \caption{The resulting age–$h_{Z}$ trend from the set of mock density distributions for a double exponential model. The left panel shows Component 1 ($h_{Z1}$), while the right panel shows Component 2 ($h_{Z2}$). The input density models had both $h_{Z1}$ and $h_{Z2}$ increasing monotonically with age in five bins (0$-$3, 3$-$5, 5$-$7, 7$-$9 and $>$9\,Gyr), with $h_{Z1}$ ranging from 0.2 to 0.6\,kpc and $h_{Z2}$ from 1.0 to 1.4\,kpc (shown by the dashed lines). We applied random errors of 30\% to the mock ages, and measured the structural parameters using our density fitting method. The fitted values with error bars are shown as points. This demonstrates the method's ability to recover the general trends of both disc components, despite the introduced age uncertainties.}
    \label{mockdata}
\end{figure*}

The results are presented in Fig.~\ref{mockdata}, which shows the recovered age-$h_{Z}$ trends for both components. Despite the introduced age uncertainties and subsequent re-binning, our method successfully recovers the general trends for both Component 1 and Component 2. The impact of age uncertainties is more pronounced for older stellar populations due to increased relative errors, leading to more significant mixing between adjacent age bins. However, this effect is mitigated to some extent by our use of wider age bins for older stars, which reduces the probability of stars being misclassified into adjacent bins. Our tests demonstrate that while a 30\% age uncertainty introduces some scatter in the derived structural parameters, it does not significantly alter the overall trends. This robustness is particularly important for older and more distant stellar populations, where age uncertainties are typically larger.

These results provide confidence in the reliability of our structural parameter measurements and their trends with age, even in the presence of substantial age uncertainties. They highlight the resilience of our methodology in recovering underlying structural trends, even when individual stars may be misclassified due to age uncertainties, suggesting that meaningful insights into the age-dependent structure of the Galactic disc can still be obtained through careful analysis of mono-age populations.

\section{Appendix}
\label{appendixB}
\begin{figure}[b]
\centering
\includegraphics[width=1.0\linewidth]{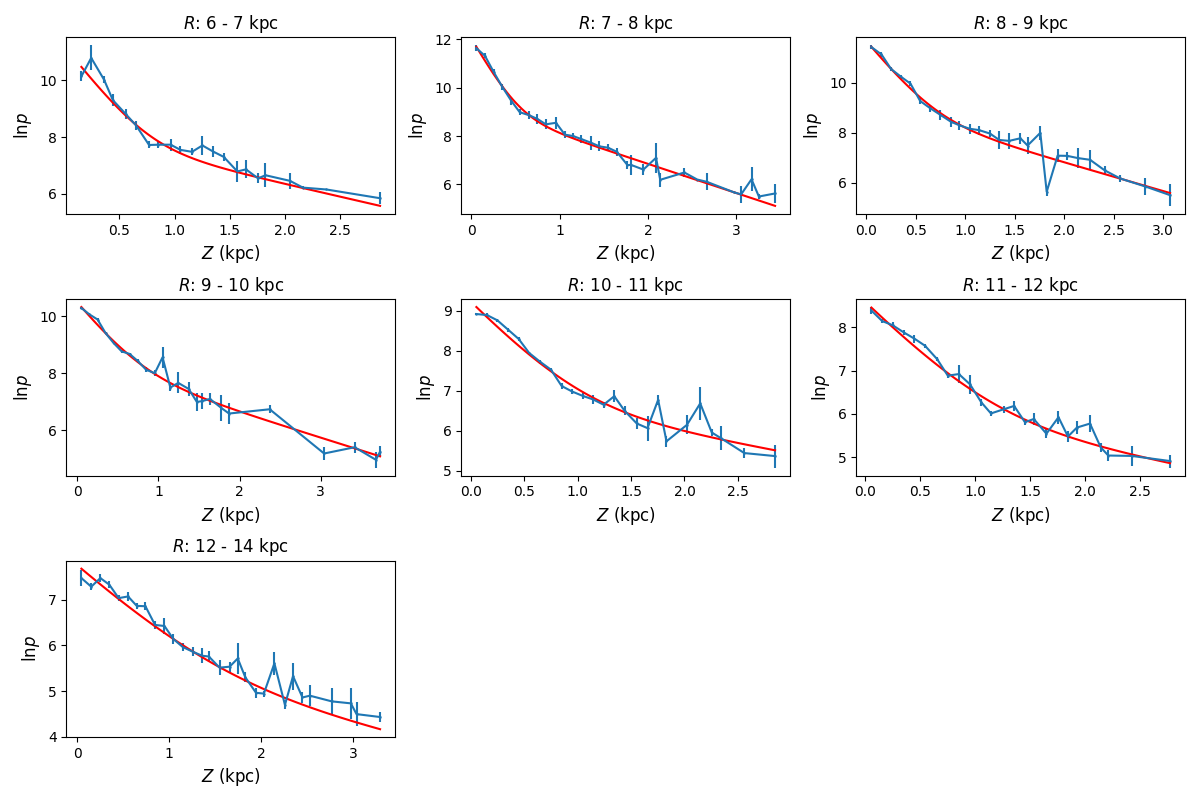}
\caption{Figures analogous to Fig.~\ref{fig3} showing the fitting results of 0$-$3\,Gyr mono-age populations. }
\label{A1}
\end{figure}

\begin{figure}
\centering
\includegraphics[width=1.0\linewidth]{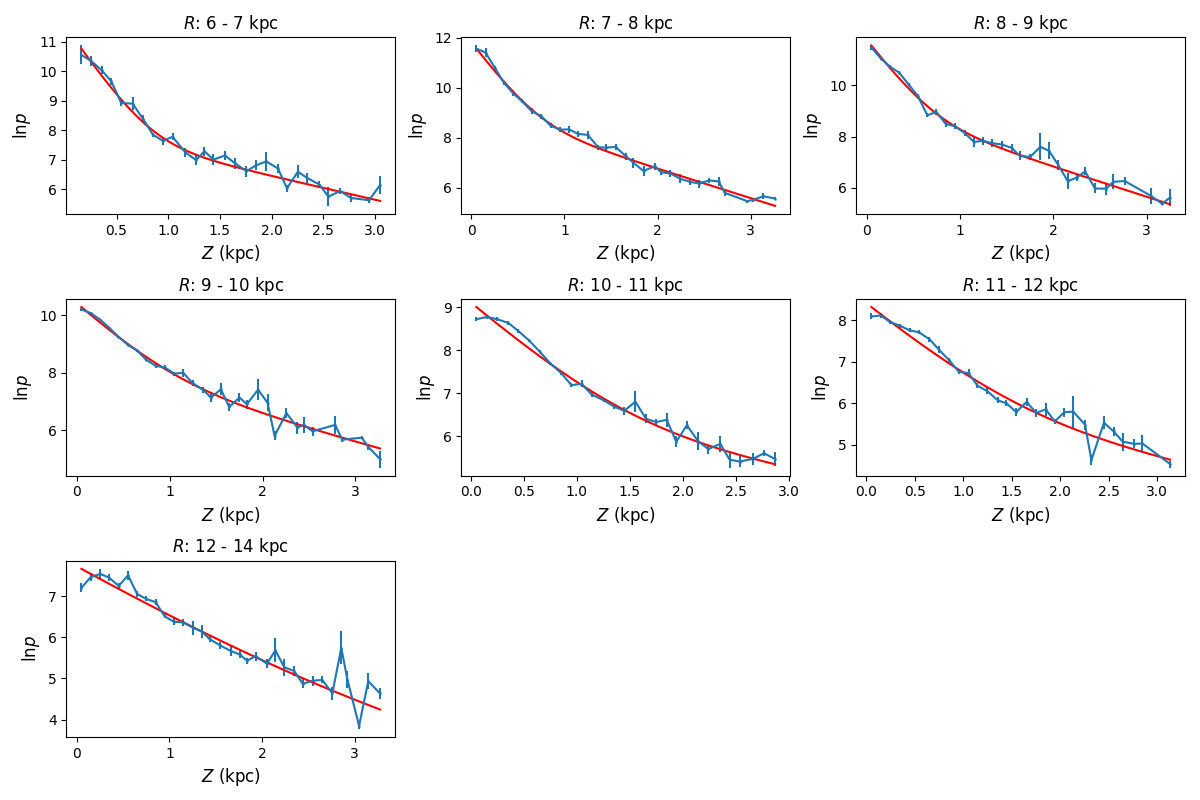}
\caption{Figures analogous to Fig.~\ref{fig3} showing the fitting results of 3$-$5\,Gyr mono-age populations. }
\label{A2}
\end{figure}

\begin{figure}
\centering
\includegraphics[width=1.0\linewidth]{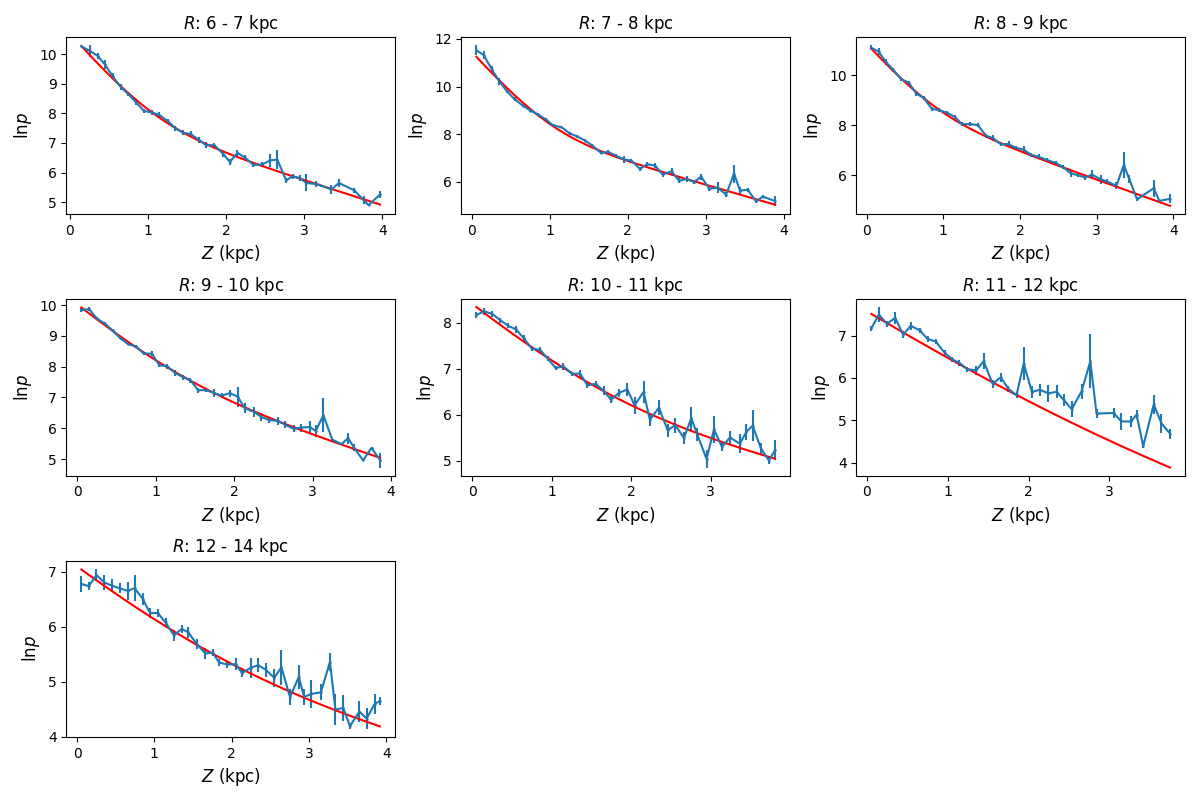}
\caption{Figures analogous to Fig.~\ref{fig3} showing the fitting results of 7$-$9\,Gyr mono-age populations. }
\label{A3}
\end{figure}

\begin{figure}
\centering
\includegraphics[width=1.0\linewidth]{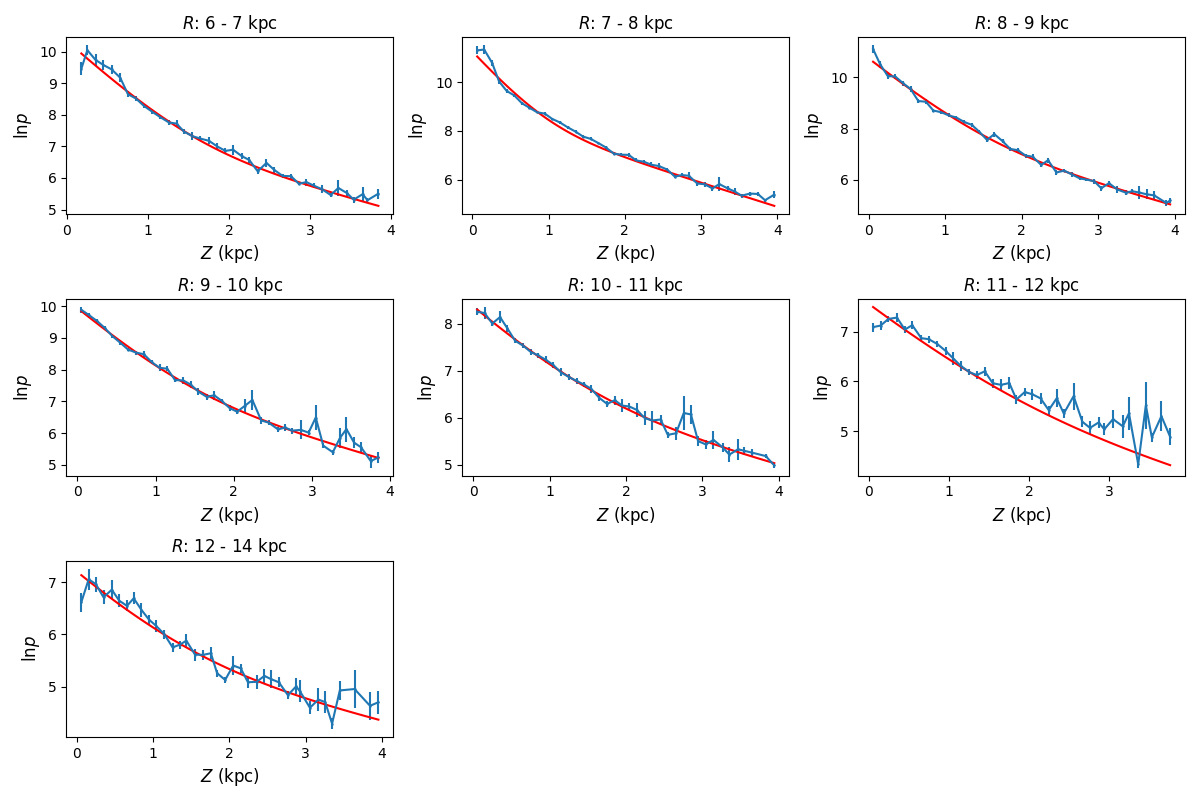}
\caption{Figures analogous to Fig.~\ref{fig3} showing the fitting results of $>$9\,Gyr mono-age populations. }
\label{A4}
\end{figure}

Fig.~\ref{A1}, \ref{A2}, \ref{A3}, and \ref{A4} present the fitting results for the mono-age populations of 0$-$3\,Gyr, 3$-$5\,Gyr, 7$-$9\,Gyr, and $>$9\,Gyr, respectively, analogous to those shown in Fig.~\ref{fig3}.

\end{document}